\documentclass[letterpaper, aps, tightenlines, nofootinbib, 11pt]{article}
\pdfoutput = 1

\usepackage{appendix}
\usepackage{setspace}
\usepackage[bookmarks = false, colorlinks = true, linkcolor = blue, citecolor = purple]{hyperref}
\usepackage{shortcuts}
\usepackage{titlesec}
\usepackage{cite}
\usepackage{amsmath}

\usepackage[margin = 2.5cm]{geometry}
\numberwithin{equation}{section}

\begin{document}

\begin{flushright}
SU-ITP 14/11
\end{flushright}

\begin{center}

\thispagestyle{empty}

\vspace*{5em}

{ \LARGE \textsc{ Chern-Simons-Ghost theories and De Sitter space \\  }}

\vspace{1cm}

{Dionysios Anninos, Raghu Mahajan, \DJ or\dj e Radi\v cevi\'c, and Edgar Shaghoulian}

\vspace{1em}

{\it Stanford Institute for Theoretical Physics and Department of Physics\\ Stanford University \\
Stanford, CA 94305-4060, USA}\\
\vspace{1em}
 \end{center}

\begin{abstract}
We explore Chern-Simons theories coupled to fundamental ghost-like matter in the large $N$ limit at 't Hooft coupling $\lambda$. These theories have been conjectured to be holographically dual to parity-violating, asymptotically dS$_4$ universes with a tower of light higher-spin fields. On $\mathbb{R}^3$, to all orders in large-$N$ perturbation theory, we show that Chern-Simons-ghost theories are related to ordinary Chern-Simons-matter theories by mapping $N \rightarrow - N$ and keeping $\lambda$ fixed. Consequently, the bosonization duality of ordinary Chern-Simons-matter theories extends to a bosonization duality of Chern-Simons-ghost theories on $\R^3$. On $S^1 \times S^2$, in the small-$S^1$ limit, neither $N \rightarrow -N$ nor bosonization hold, as we show by extensively studying large-$N$ saddles of the theories with both ghost and ordinary matter. The partition functions we compute along the way can be viewed as pieces of the late-time Hartle-Hawking wavefunction for the bulk dS$_4$ gravity theories.

\end{abstract}

\pagebreak
\setcounter{page}{1}

{\small
\tableofcontents
}

\section{Introduction and summary}

Chern-Simons-matter (CS-matter or CSM) theories consist of a Chern-Simons gauge field coupled to a matter sector. It has been proposed that the large $N$ limit of certain field theories of this type is holographically dual to a class of exotic gravity theories with a tower of light higher-spin particles (Vasiliev theories of gravity). The duality has been proposed for both asymptotically AdS$_4$ \cite{Vasiliev:1992av, Vasiliev:1995dn, Vasiliev:1999ba, Vasiliev:2003ev, Klebanov:2002ja, Sezgin:2002rt, Giombi:2012ms, Petkou:2003zz, Sezgin:2002ru, Sezgin:2003pt, Giombi:2009wh, Giombi:2010vg, Girardello:2002pp, Giombi:2011ya, Chang:2012kt, Das:2003vw, Koch:2010cy, Douglas:2010rc, Jevicki:2011ss} and asymptotically dS$_4$ backgrounds \cite{Anninos:2011ui, Ng:2012xp, Anninos:2012ft, Anninos:2013rza, Banerjee:2013mca, Chang:2013afa, Das:2012dt}. 

It is of note that such conjectured boundary duals of de Sitter space contain ghosts --- matter fields with ``wrong'' statistics such as commuting fermions or anticommuting bosons. Such CS-ghost theories violate the spin-statistics connection and are non-unitary in the Euclidean sense: they are not reflection-positive.
Though the study of non-unitary statistical field theories may sound unusual, it is worth recalling that in statistical field theories the unitarity condition need not be a physical condition. Indeed \cite{difrancesco}, \newline\newline
{{\it ``Statistical models of hard objects always admit critical continuum descriptions with non-unitary conformal field theories. Moreover, many other physical systems such as polymers in two dimensions have phases described by non-unitary minimal models."}}
\newline\newline
Furthermore, if de Sitter space is to have a global holographic description as envisioned in \cite{Witten:2001kn, Strominger:2001pn, Maldacena:2002vr}, the dual theories living on late-time Euclidean surfaces will be non-unitary. Guided by these ideas, in this paper we perform a detailed study of certain properties of non-unitary CS-ghost models that may shed further light on the dS/CFT correspondence.

\subsection*{Structure of this paper}

\begin{figure}[t!]
  \centering
  \includegraphics[width=1\textwidth]{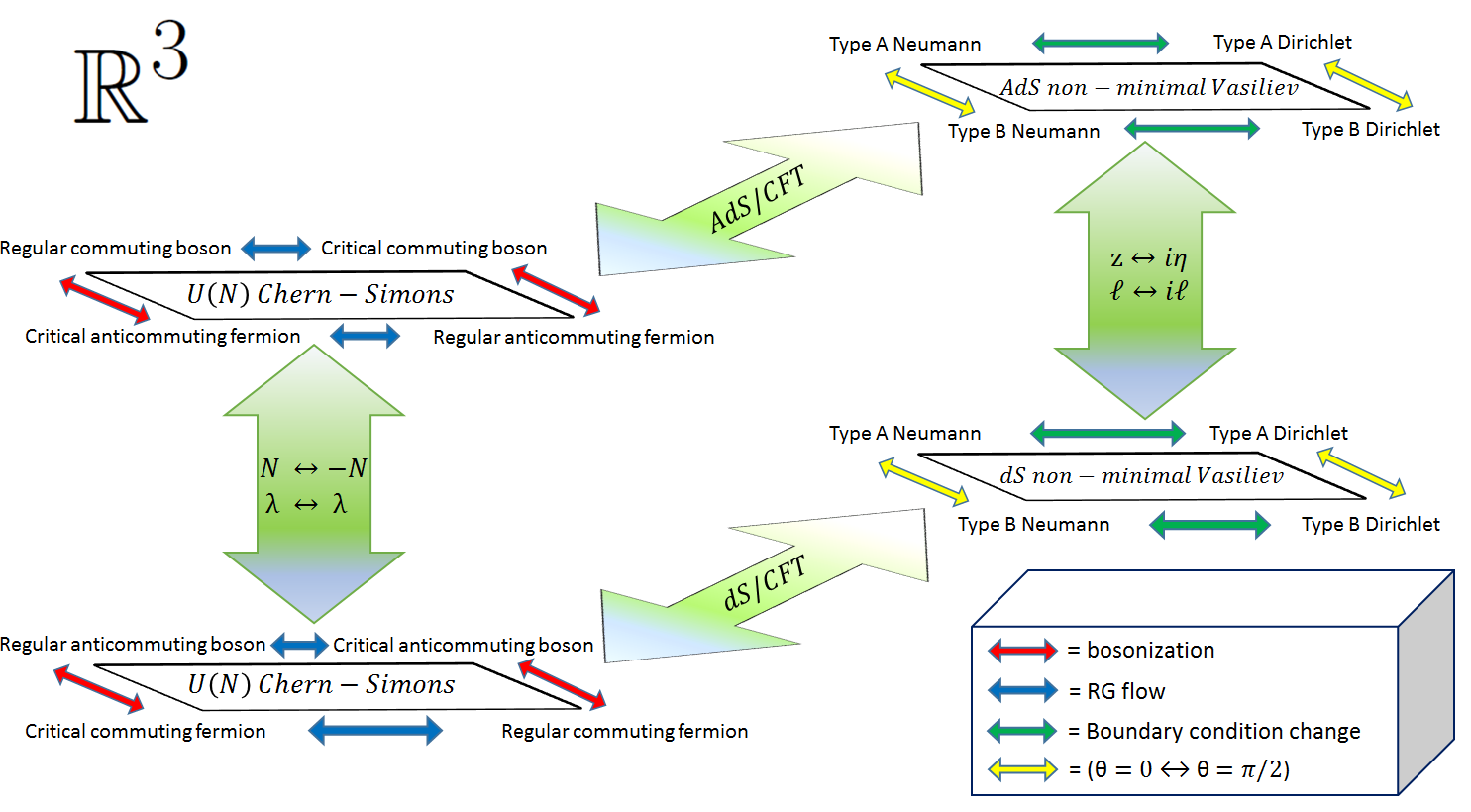}
  \caption{\small (Color online) All currently known duality maps between $U(N)$ Chern-Simons theories on $\mathbb{R}^3$ and parity-violating all-spin Vasiliev theories in de Sitter and anti-de Sitter, including the results in this paper. It should be noted that the bosonization maps hold along the RG flow and not just at the endpoints where the arrows are displayed. Similarly, for the bulk theories one can get from Type A with mixed boundary conditions to Type B with mixed boundary conditions by changing the parity-violating phase (although not in the simple way stated in the legend). An analogous set of dualities exists for the even-spin Vasiliev theories.}
  \label{fig arrowsdiag}
\end{figure}

In the remainder of this Introduction we further motivate our study by discussing general aspects of dS/CFT and some features of CS-matter theories. In Sec.~\ref{sec CSG}, we discuss the realm of possible CS-ghost models and introduce the actions of all the models that we will be studying in greater detail. In Sec.~\ref{sec T0}, we study the chosen models on $\R^3$. Using two different approaches --- perturbative computations in Landau gauge and path-integral methods in the light-cone gauge --- we demonstrate that the $N \rightarrow -N$ map works and that, by proxy, all correlators of relevant fermionic ghost theories can be ``bosonized'' into correlators of corresponding bosonic ghost theories. These results fit into a web of dualities encapsulated by Fig.~\ref{fig arrowsdiag}. Along the way, the two-loop $\beta$-functions for the regular bosonic ghost theories are computed and shown to vanish exactly at infinite $N$. In Sec.~\ref{sec Thigh}, we analyze these models on $S^1 \times S^2$ in the limit of a very small $S^1$. We find that both $N \rightarrow -N$ and bosonization are broken, but along the way we extend the previously available small-$S^1$ analysis to find new saddle points and small-$S^1$ Bardeen-Moshe-Bander phase transitions \cite{Bardeen:1983rv} for theories with both ordinary and ghost matter. We finish with some conclusions and an overview of possible future research directions.
\newline\newline As an introductory exercise, and in order to place the theories we consider into a larger framework, we now proceed to examine the general roles of gauge fields in holography.

\subsection*{Gauge fields in AdS/CFT}

Gauge fields play a very important role in our understanding of the AdS/CFT correspondence. In the simplest examples we know of, where the CFT is given by an IR fixed point of the world-volume theory on a stack of $N$ D-branes, the gauge fields are nothing more than the order $N^2$ massless modes of the strings interconnecting the D-branes.

As a first example, consider the four-dimensional $\mathcal{N}=4$ SYM with gauge group $SU(N)$. At strong `t Hooft coupling, this theory is dual to AdS$_5 \times S^5$ in type IIB string theory. The matter content of the theory consists of a vector multiplet containing the gauge field $A^{ij}_\mu$, with $i = 1,\ldots,N$, six real scalars $\Phi^I$, with $I=1,\ldots,6$, and the fermionic partners. All the fields transform in the adjoint. The bosonic part of the action is given schematically by
\begin{equation}
S\_{SYM} = \int \d^4 x \; \Tr \left( \frac{1}{2 g^2} \, F_{\mu\nu} F^{\mu\nu}  + D_\mu \Phi^I D^\mu \Phi^I + \frac{g^2}{2} \, \left[ \Phi^I, \Phi^J \right]^2 \right),
\end{equation}
where $F_{\mu\nu}$ is the $SU(N)$ field strength and $D_\mu$ is the gauge-covariant derivative. The theory contains a vast collection of single-trace (gauge-invariant) operators, the large majority of which have a conformal weight that is not protected by supersymmetry as the dimensionless 't Hooft coupling $\lambda = g^2 N$ is taken to be large.\footnote{A simple example of single trace operators with {\it protected} weights is the collection of chiral primaries: $\O^{I_1\ldots I_n} = \Tr \; \Phi^{I_1} \Phi^{I_2} \ldots \Phi^{I_n} + \ldots$, where terms are added to make $\O^{I_1\ldots I_n}$ traceless and symmetric and $n<N$. These are dual to Kaluza-Klein modes on the $S^5$ in the bulk.} This collection of operators manifests itself in the bulk as a collection of heavy closed string states whose mass $m_s$ in AdS units goes as some power of $\lambda$.
Specifically, the conformal weight of an operator dual to a bulk string state is $\Delta_s(\lambda) \sim m_s \ell\_{AdS} \sim \lambda^{1/4}$. At weak coupling however, when $\lambda$ is small, the dependence on $\lambda$ becomes a perturbative effect and the collection of operators no longer decouples from the light part of the spectrum.

In three boundary dimensions, the Yang-Mills gauge coupling acquires dimensions and grows large in the deep IR. Fixed points have been identified for ordinary gauge theories coupled to an $N$-component field in the large $N$ limit \cite{Appelquist:1988sr,Appelquist:1981sf}. However, one can also construct (super)-conformal field theories containing a CS gauge field (with no ordinary two-derivative kinetic term for the gauge field) coupled to matter. A simple example is given by the ABJM model \cite{Aharony:2008ug}, a $U(N)_{k} \times U(N)_{-k}$ CS theory with levels $k$ and $-k$ for each gauge group, and with $\mathcal{N}=6$ supersymmetry. In addition to the vector multiplet, the theory contains a bifundamental chiral multiplet $\Phi^{IJ}$, where the $I$ ($J$) are fundamental (anti-fundamental) indices for each $U(N)$ gauge group. At infinite $k$, the theory is dual to AdS$_4 \times \mathbb{C}\text{P}^3$ in type IIA string theory. By turning on a magnetic $B$-flux on the $\mathbb{C}\text{P}^3$, the gauge group can be generalized to a $U(N)_k \times U(M)_{-k}$ group, and the resulting theory is known as the ABJ theory \cite{abj}. Notice that since the gauge field is no longer dynamical, there are no local single-trace operators of the form $\O = \Tr F \, F \ldots F$. There is, however, a large collection of local gauge-invariant single-trace operators of the form $\O = \Tr \; \Phi^{IJ} \Phi^{JK} \Phi^{KL} \ldots \Phi^{ZI}$, where by the trace symbol we simply mean a sum over all repeated bifundamental indices. Again, the conformal weight of a vast collection of such operators is not protected and can grow with an increasing 't Hooft coupling. These are dual again to heavy bulk string states.

In the limit where $N \gg M \sim \O(1)$, the structure of single-trace operators changes significantly \cite{Chang:2012kt}. The above strings of operators are better understood as multitrace operators of a single $U(N)_{k}$ gauge theory, and the single-trace operators become simple bilinear pairs of the $\Phi^{IJ}$ with the $U(N)$ index summed over. This is the vector limit of the theory, which has been argued extensively to be dual to a higher-spin, parity-violating, bulk theory \cite{Giombi:2011kc,Aharony:2011jz}. Gauge fields in these theories play a rather different (perhaps somewhat more innocent) role than they did in the four-dimensional case discussed above. They do not give rise to a large collection of single-trace local operators dual to bulk fields. Instead they are there to guarantee the gauge invariance of the theory and constrain the structure of the actual set of local operators.

Finally, it is important to point out that unlike the above two examples of theories with bulk duals, there are many cases where the CFT has no weakly-coupled Lagrangian description nor a marginal coupling that can be continuously taken to zero. In fact, the very intuition of the $N^2$ gauge fields being the light string modes on the stack of $N$ D-branes breaks down in AdS$_7 \times S^4$ vacua of M-theory appearing in the near horizon of a stack of M5-branes. Indeed, eleven-dimensional M-theory does not have strings as fundamental degrees of freedom, and it is most likely that we cannot understand such CFTs as conventional gauge theories in any sense.\footnote{More generally, in the spirit of the bootstrap program (see for example \cite{Polyakov:1974gs,Rattazzi:2008pe}), a CFT need not be defined by a Lagrangian but rather as a collection of local operators, OPE coefficients, and anomaly coefficients, and as such it is subject to consistency requirements such as crossing symmetries, unitarity, and so on.}

\subsection*{Gauge fields in dS/CFT}

The goal of dS/CFT is to understand whether there exist CFTs that describe (globally) an asymptotically de Sitter universe (see \cite{Spradlin:2001pw,Anninos:2012qw} for an overview of issues regarding quantum de Sitter space). Though little is known about such CFTs, there are some hints coming from simple bulk computations. For instance, the conformal weight of an operator dual to a bulk scalar field of mass $m$ in four dimensions is given by $\Delta = \frac32 \pm \sqrt{\frac94-m^2 \ell^2\_{dS}}$, and so a tachyonic field with $m^2 \ell^2\_{dS} \ll -1$ would be dual to an operator with large real conformal weight. Thus, in constructing (meta-)stable de Sitter spacetimes out of CFTs that contain weakly-coupled dynamical gauge fields, one has to confront the fact that the tower of single-trace operators $\O = \Tr F \, F \ldots F$ with increasing number of $F$'s (and thus increasingly large real conformal weights) will be dual to a tower of increasingly tachyonic bulk fields. Though such a ``tachyonic catastrophe'' most clearly occurs for weakly coupled gauge theories, it is reasonable that it will occur at strong coupling as well. This  suggests that the CFTs related to asymptotically stable de Sitter universes are of a different kind than standard gauge theories (see \cite{Banerjee:2013mca} for a related discussion), and one is lead to consider: (i) CFTs that contain non-dynamical gauge fields, such as the three-dimensional CSM theories with ghost matter, or (ii) more generally CFTs that do not have a canonical weakly-coupled Lagrangian description.\footnote{See however \cite{Bzowski:2012ih, McFadden:2009fg} for a discussion of extracting CMB correlators from more general putative holographic models.}

The first case has led to some progress in the study of a class of CFTs which are dual to four-dimensional asymptotically de Sitter universes containing an infinite tower of higher-spin fields. In order to avoid the ``tachyonic catastrophe'' in the bulk, one has to further go to the deep ABJ limit in the boundary, where the theories are vector-like rather than matrix-like and the single-trace operators are simply bilinear in the matter fields. (This field-theoretic limit was analyzed in some detail in \cite{Banerjee:2013nca}.) The CFTs of interest in this context contain ghost-like matter fields transforming in the fundamental or antifundamental representation of a particular group, such as anticommuting scalar fields or commuting fermions. It has further been argued that they can be coupled to an {\it ordinary} \cite{Anninos:2013rza,Banerjee:2013mca,Chang:2013afa} CS gauge theory, thereby giving rise to a family of CFTs that are dual to a parity-violating family of bulk higher-spin dS theories. These CS-ghost theories will be the main focus of the present work.

In the second case, the task may amount to understanding how to extend the bootstrap problem to the non-unitary CFTs dual to asymptotically de Sitter bulk theories. The main challenge, as compared to the original bootstrap program, is the abandonment of unitarity as a requirement. (For two-dimensional CFTs, however, one also has additional constraints due to modular invariance.) Instead, what comes into play is the condition of bulk unitarity, which at tree level is simply the requirement that the bulk Lagrangian give rise to a Hermitian Hamiltonian.\footnote{One might also require Gaussian suppression of the bulk Hartle-Hawking wavefunction \cite{Hartle:1983ai,Hertog:2011ky} (at least for deformations near the de Sitter vacuum) as a condition. This can be accomplished by ensuring that the two-point function of those operators dual to bulk fields is negative definite; see \eqref{pertds}.} Consider for instance the action of a massless scalar field with a quartic interaction about a fixed planar four-dimensional de Sitter background (with metric $\d s^2 = {\ell^2\_{dS}} \left( -\d\eta^2 + \d\vec{x}^2 \right)/\eta^2$ and $\eta < 0$):
\begin{equation}
S = \frac{\ell^2\_{dS}}{2}\int \d^3 \vec{x} \, \frac{\d\eta}{\eta^4}  \left(\eta^2 \, \partial_\eta \phi \, \partial_\eta \phi - \eta^2 \, \partial_i \phi \, \partial_i \phi - \frac{\ell_{\text{dS}}^2\lambda_4}{4!} \, \phi^4 \right).\label{quarticaction}
\end{equation}
The index $i$ ranges over the spatial indices in the bulk. A Hermitian Hamiltonian for $\phi$ requires $\lambda_4 \in \mathbb{R}$ and stability requires $\lambda_4 \ge 0$. The late-time tree-level Bunch-Davies wavefunctional of the scalar as a function of the late-time profile $\varphi(\vec{x}) = \lim_{\eta \rar 0} \phi(\vec{x},\eta)$ to $\O(\lambda)$ is given by:
\begin{equation}
\lim_{\eta\rar 0}\log |\Psi\_{BD}[\varphi;\eta] | = - \ell^2\_{dS}\! \int \d^3 {\vec{x}}\, \d^3 {\vec{y}} \left( \frac{6}{\pi^2} \, \frac{ \varphi(\vec{x}) \varphi(\vec{y})}{|\vec{x}-\vec{y}|^6} + \lambda_4 \, \ell\_{dS}^2 \! \int\!  \d^3 {\vec{v}} \,\d^3 {\vec{w}} \;  \varphi(\vec{x}) \varphi(\vec{y})  \varphi(\vec{v}) \varphi(\vec{w}) \, \mathcal{I} \right),\label{quarticmess}
\end{equation}
where
\begin{equation}
\mathcal{I} =  \int \frac{\d\eta \, \d^3\vec{q}}{\eta^4} K(\eta,\vec{x}) K(\eta,\vec{y}) K(\eta,\vec{v}) K(\eta,\vec{w})~, \quad K(\eta,\vec{w}) \equiv  \frac{6}{\pi^2}\, \left(  \frac{i \eta}{\eta^2 - |\vec{w}-\vec{q}|^2} \right)^3~.
\end{equation}
The $i \, \varepsilon$ prescription for the $\eta$-singularities in the $\mathcal{I}$ integral is such that the integral is evaluated on the slightly deformed contour defined by the $\eta + i\epsilon$ axis. (The calculation can also be viewed as an analytic continuation from Euclidean AdS$_4$ \cite{Maldacena:2002vr}.) 
According to the dS/CFT correspondence, $\lim_{\eta\rar 0}\Psi\_{BD}[\varphi(\vec{x}),\eta]$ is given by the partition function of a putative dual CFT as a function of the source $\varphi$ for the operator $\O_\phi$ dual to $\phi$:
\begin{equation}
\lim_{\eta\rar0}\;\log \Psi\_{BD}[\varphi(\vec{x}),\eta] = \sum_{n=1}^{\infty} \frac{1}{n !} \int  \d^3 {\vec{x}}_1 \ldots \int \d^3 {\vec{x}}_n  \; \varphi(\vec{x}_1) \ldots \varphi(\vec{x}_n) \; \langle \mathcal{O}_\phi(\vec{x}_1) \ldots  \mathcal{O}_\phi(\vec{x}_n) \rangle\_{CFT}.\label{pertds}
\end{equation}
From the CFT point of view, hermiticity of the bulk Hamiltonian is thus related to reality properties in the correlation functions of the CFT. 

For the field theory \eqref{quarticaction}, odd terms in the sum will vanish by a $\phi\rightarrow -\phi$ symmetry. In this case, knowing the precise bulk action gives us the precise boundary correlators. One might imagine constructing a machine which takes as input complete CFT data and outputs whether the CFT can be dual to a bulk de Sitter theory. A minimal task for this machine is to construct the bulk late-time correlation functions and check that they are consistent with a late-time bulk Hamiltonian which is Hermitian.

We have illustrated how bulk unitarity constrains the boundary field theory in the simple case of a scalar field with quartic self-coupling, but the connection holds generally. An interesting example is a bulk theory admitting Einstein-like gravitational interactions, such that tree-level correlators between other fields and the graviton are suppressed. Boundary correlators involving  the tree-level graviton correspond to tree-level correlators involving only the stress tensor. These will be real (up to local terms) for a four-dimensional bulk, so long as the coefficient of the two-point function is normalized to be negative definite. Thus, even at the basic tree level we are confronted with the challenge of extracting from the CFT the reality properties of the bulk couplings. We will leave a discussion of this approach to other work.

Finally, we would like to remark on the idea of defining the CFT via an analytic continuation of some known CFT dual to an AdS theory. The typical issue that arises from the bulk theory, particularly in the case of supergravity, is that many couplings depend on the bulk cosmological constant and the continuation can lead to ghosts and tachyons. More abstractly, a CFT is defined by the collection of conformal weights, spins, and OPE coefficients $\{\Delta_i,s_i,c_{ijk} \}$ obeying the OPE associativity at the level of four-point functions. Assume now that we have a consistent set of data $\{\Delta_i,s_i,c_{ijk} \}$ defining some AdS theory. It is unclear whether there exists an unambiguous procedure to analytically continue this data to some dS theory.\footnote{For example, imagine a mass $m_A$ bulk scalar field in AdS$_{d+1}$ (with AdS length $\ell_A$) dual to a scalar operator with weight $\Delta_A(\Delta_A-d) = m_A^2 \ell_A^2$. It might seem natural the continued theory have an operator $\Delta_D$ such that $\Delta_D(\Delta_D-d) = - m_A^2 \ell\_{dS}^2$ due to the continuation $\ell_A \to i \ell\_{dS}$. However, this is not what occurs in higher-spin examples of de Sitter holography, where we must also continue $m_A^2 \to - m_A^2$, thus leaving the conformal weights unchanged.}

\subsection*{Mappings between different CS-matter theories}

The interest in CSM holography --- both AdS and dS --- has sparked a renewed interest in CSM theories in their own right. Early investigations of such models have been largely fueled by the novel understanding of ubiquitous properties of CS theory \cite{Witten:1988hf, Elitzur:1989nr} and by the interest in the fractional quantum Hall effect \cite{Zhang:1988wy, Wen:1990zza, Zee:1996fe}. Among these early advances, the one of greatest relevance for our work is the discovery that $SU(N)_k$ CS coupled to external bosonic sources (i.e. level-$k$ CS theory with bosonic Wilson lines in the fundamental representation of $SU(N)$) is dual via level-rank duality to the $SU(k)_{-k - N\sgn k}$ CS coupled to external fermionic sources (i.e.~to CS theory with fermionic Wilson lines, also in the fundamental representation). Recent years have seen the emergence of the analogous ``bosonization'' statement for CS theories coupled to quantum matter \cite{Aharony:2011jz, Aharony:2012nh, Aharony:2012ns, Maldacena:2011jn, Maldacena:2012sf, Jain:2012qi, Jain:2013gza, Jain:2013py, Takimi:2013zca, Yokoyama:2012fa, GurAri:2012is, Shenker:2011zf, Banerjee:2012gh, Giombi:2011kc}.\footnote{Most statements and checks were made in the large $N$ limit, but the arguments of \cite{Jain:2013py} should apply to theories with any $k$ and $N$.} Moreover, various technical tools have been developed or brought up-to-date in order to provide powerful computational handles on large $N$ limits of CSM theories at zero temperature \cite{Giombi:2011kc, Aharony:2012nh, GurAri:2012is, Jain:2013gza, Maldacena:2011jn, Maldacena:2012sf, Yokoyama:2012fa, Aharony:2011jz}, at high temperatures \cite{Jain:2012qi, Takimi:2013zca, Shenker:2011zf,Aharony:2012ns,Jain:2013py}, and even on non-trivial background topologies \cite{Banerjee:2012gh, Radicevic:2012in, Banerjee:2013mca}. Rich phase diagrams and duality structures have been discovered, and with this machinery at hand, it is natural to ask how the models conjectured to be dual to dS spaces fit into this intricate framework.

As stressed in the previous subsection, the CFTs of interest are theories in which a CS sector is coupled to a large number $N$ of ``ghost'' fields, i.e.~fields which violate the spin-statistics connection. These CS-ghost theories can be realized either via anticommuting bosons or via commuting Dirac fermions. In this note we initiate a systematic taxonomy of these theories. We identify four CS-ghost models which, on an $\R^3$ manifold, can be \emph{mapped} to models of CS coupled to ordinary matter:
\begin{itemize}
  \item The $Sp(2N,\mathbb{R})$ 
  CS theory coupled to $2N$ anticommuting bosons, mapped by $N \rightarrow - N$ to the ordinary $O(2N)$ CS-boson theory.
  \item The $Sp(2N,\mathbb{R})$ 
  CS theory coupled to $2N$ commuting fermions, mapped by $N\rightarrow -N$ to the ordinary $O(2N)$ CS-fermion theory.
  \item The $U(N)$ CS theory coupled to $N$ anticommuting bosons, mapped by $N \rightarrow -N$ to the ordinary $U(N)$ CS-boson theory.
  \item The $U(N)$ CS theory coupled to $N$ commuting fermions, mapped by $N \rightarrow -N$ to the ordinary $U(N)$ CS-fermion theory.
\end{itemize}
We explain the nomenclature in Sec.~\ref{sec CSG} and in Appendix \ref{Appendix SpN}.
See Fig.~\ref{fig arrowsdiag} for a pictorial depiction of the various theories.
In the first two bullet points above, when the theory is put on manifolds with nontrivial topology, there are subtleties regarding the noncompact nature of the group $Sp(2N,\mathbb{R})$. We expand on this in greater detail in Appendix \ref{Appendix SpN}, where we consider the ``restricted $USp(N)$" CS theory with a compact gauge group.

These mappings between field theories were originally conjectured in \cite{Anninos:2013rza, Chang:2013afa}, with evidence for $U(N)$ at infinite $N$ provided in \cite{Chang:2013afa}. The case of a pure Maxwell gauge theory case was proven in \cite{Mkrtchian:1981bb}. We will present detailed computations and arguments supporting the general case for the four models listed above to \emph{all} orders in $N$ and at all values of couplings. 
 We will also show that the $N\rightarrow -N$ dualities in the two $U(N)$ models break on nontrivial topology (on $S^1 \times S^2$ in the limit of small $S^1$) at leading order in $N$. 

Another natural duality to investigate is the bosonization map between bosonic and fermionic CS-ghost models. This duality is particularly interesting because there exists bulk evidence that it should hold, as the dS higher-spin gravities dual to the bosonic and the fermionic theory have been shown to be dual to each other \cite{Chang:2013afa}. Using the $N \rightarrow -N$ map and the fact that this bosonization was demonstrated for ordinary CS-matter theories \cite{Aharony:2012ns, Aharony:2012nh}, we conclude that bosonization holds for $U(N)$ and 
$Sp(2N,\mathbb{R})$ CS-ghost theories on $\R^3$. This is a novel example of bosonization for non-unitary theories. We then study the $U(N)$ CS-ghost theories on $S^1 \times S^2$ in the limit of small $S^1$, where we may no longer use the $N \rightarrow - N$ map, and we carefully examine the large-$N$ saddle-point structure. We find that bosonization fails to hold, with the dominant saddle in the bosonic theory being mapped to a sub-dominant saddle which is not expected to lie on the steepest descent contour in the fermion theory.


\section{CS-ghost theories}\label{sec CSG}
We begin overviewing the family of theories that can be obtained by gauging a global symmetry of a vectorlike ghost model and coupling it to a a CS gauge sector. We are particularly interested in theories whose spectra of conserved currents correspond to even-spin (minimal) and all-spin (non-minimal) spectra of Vasiliev fields in four dimensions.

Ungauged vector models with $N$ anticommuting bosons are highly constrained by the Grassmannian nature of the fields. They can have four simple global symmetries that are distinct at large $N$: $U(N)$, $USp(N)$, $Sp(N, \R)$, and $Sp(N, \C)$  (see Appendix \ref{Appendix SpN} for definitions and conventions). The first two groups are compact and can be gauged without issues. The latter two are non-compact, and gauging them can lead to various non-perturbative divergences \cite{BarNatan:1991rn}; moreover, they can only be defined for even $N$.
The $U(N)$ ghost model has precisely the right singlet content in order to be dual to the all-spin Vasiliev theory in dS$_4$. On the other hand, the singlet content of the $USp(N)$ ghost model contains currents of odd spin, making it ineligible to be dual to the minimal (even-spin) Vasiliev gravity.
By restricting this model to just the ``real'' fields, a theory with a single conserved current of each even spin is obtained.
On $\R^3$, this restricted $USp(N)$ ghost model can be viewed as an analytic continuation of the $Sp(2N, \R)$ model; they have the same conserved current content, and we call the resulting theory on $\mathbb{R}^3$ the $Sp(2N,\mathbb{R})$ ghost model.

Vector models with $N$ commuting fermions behave similarly. The $U(N)$ commuting fermion theory has a single conserved current of each integer spin, while the $Sp(2N,\mathbb{R})$ commuting fermion has a single current of each even spin.


For each global symmetry group and choice of ghost spin, there exist two conformal theories that can be coupled to the CS sector. For bosonic ghosts, these are the regular theory (the free theory deformed by a marginal sextic term) and the critical theory (the Wilson-Fisher fixed point, or the IR limit of the free theory deformed by a relevant quartic term). For fermionic ghosts, we again have a regular theory (free fermions, no marginal deformations) and a critical theory (the Gross-Neveu fixed point with a marginal cubic term, or the theory whose deformation by a quartic term flows to the free theory in the IR). See Fig.~\ref{fig arrowsdiag} for a pictorial depiction of the various theories and the mappings and dualities between them.

The path integrals of all theories of interest can now be written explicitly. Let us start with the two $U(N)$ bosonic models with an $N$-component complex matter field $\chi_i$ with Hermitian conjugate $\bar \chi_i$. The critical theory partition function is a Legendre transform of the regular theory partition function, so we may analyze both simultaneously by computing
\bel{\label{def ZB}
  Z_B = \int [\d A\, \d\chi\, \d\sigma] e^{-S\_{CS} - S\_{B} - \int \d^3 x\, \sigma \bar\chi_i \chi_i + \frac N{2\lambda_4^b} \int \d^3 x\, \sigma^2},
}
with
\bel{\label{def SCS}
  S\_{CS} = -\frac{ik}{8\pi} \int \d^3 x\, \epsilon^{\mu\nu\rho} \left(A_\mu^a \del_\nu A_\rho^a + \frac13 f^{abc} A_\mu^a A_\nu^b A_\rho^c \right)
}
and
\bel{\label{def SB}
  S\_B = \int \d^3x \left( (D_\mu \chi)_i\+ (D^\mu \chi)_i + N \frac{\lambda_6^b}{3!}\left(\frac{\bar\chi_i \chi_i}N\right)^3\right),\quad D_\mu \equiv \del_\mu + A_\mu.
}
The generators $T^a$ of the $U(N)$ group are all taken to be anti-Hermitian, and a general group element of $U(N)$ is given by $e^{\theta^a T^a}$. Upon integrating out the Hubbard-Stratonovich auxiliary field $\sigma$, we obtain the critical $(\bar{\chi}_i \, \chi_i)^2$ interaction. The regular bosonic theory possesses a marginal, six-point deformation which has been included in the above action. This deformation does not affect the critical theory, as will become clear from our final results.

The path integral over $\sigma$ in \eqref{def ZB} runs over the real axis, and $\lambda_4^b$ is the quartic coupling (by dimensional analysis, it is proportional to the inverse energy scale); the flow from $\lambda_4^b = 0$ to $\lambda_4^b = \infty$ represents the flow from the UV to the IR, i.e.~from the free to the critical theory.\footnote{Ref.~\cite{Anninos:2013rza} has analyzed such theories for $\lambda_4^b \in \C$, but in this work this coupling is purely real.} In order to avoid clutter, when working with the regular theory we will just set $\sigma = 0$ and discard this path integration. When working with the critical theory, we will send $\lambda_4^b\rar \infty$ at the outset; one can verify that this is equivalent to keeping $\lambda_4^b$ finite, integrating over $\sigma$, 
and then letting $\lambda_4^b \rar \infty$.

The path integral relevant to the two $Sp(2N, \R)$ bosonic theories can be obtained from the $U(N)$ one \eqref{def ZB} by substituting $\bar\chi_i \chi_i \rightarrow \frac12\Omega_{ij} \chi_i \chi_j$ in \eqref{def SB} and increasing the range of matter indices to $2N$. It is also necessary to substitute $A^a_\mu \del_\nu A^a_\rho$ in \eqref{def SCS} with $\eta^{ab} A_\mu^a \del_\nu A_\rho^b$, where $\eta^{ab}$ is the Cartan-Killing metric of $Sp(2N, \R)$ (see Appendix \ref{Appendix SpN}). Of course, the structure constants $f^{abc}$ must be changed appropriately, as well.

Finally, let us address the $U(N)$ ghost fermion theory. The path integral that captures both the regular and critical theory is
\bel{\label{def ZF}
  Z_F = \int[\d A\, \d \xi\, \d \sigma] e^{-S\_{CS} - S\_F - \int \d^3x\, \sigma \bar\xi_i \xi_i + \frac{N}{3!} \lambda_6^f \int\d^3x\,\sigma^3},
}
The CS term is the same as before (eq.~\eqref{def SCS}), and the $N$ Dirac (two-component) fermions $\xi_i$ have action
\bel{\label{def SF}
  S\_F = \int\d^3 x\, \bar\xi_i \gamma^\mu (D_\mu \xi)_i.
}
The $\gamma$ matrices are just the three Pauli matrices. As before, the regular fermion theory is obtained by setting $\sigma = 0$ and discarding that path integral.

\section{CS-ghost theories on $\R^3$}\label{sec T0}

The objects of principal interest in studying dualities on $\R^3$ are correlation functions of the gauge-invariant ``single-trace'' bilinears $\O_{\mu_1\ldots \mu_n} \sim \Omega_{ij} \chi_i D_{\mu_1} \ldots  D_{\mu_n} \chi_j$ or $\O_{\mu_1\ldots\mu_n} \sim \bar\chi_i D_{\mu_1} \ldots D_{\mu_n} \chi_i$ (and the analogous fermionic operators). By choosing the right combination of derivatives, all these operators can be made transverse and traceless in the spatial indices, and they correspond to higher-spin conserved currents. Their expectation values, as well as any other correlators of interest, may be computed as usual by including source terms in the partition functions \eqref{def ZB} and \eqref{def ZF} and differentiating these source-dependent functionals $Z_{B/F}[J]$. Alternatively, these correlators may be computed by straightforward perturbation theory. We take both routes: first we perturbatively calculate the $\beta$-functions and demonstrate the conformality properties of the theories in question, and then we calculate the generating functional by directly integrating out matter. In both cases we find that regular and ghost models are related by $N \rightarrow -N$ and $\lambda$ fixed, as was anticipated in \cite{Chang:2013afa}.

\subsection{Perturbative approach in Landau gauge}
\label{section: landau}

\begin{figure}
\begin{center}
\includegraphics[width=\textwidth]{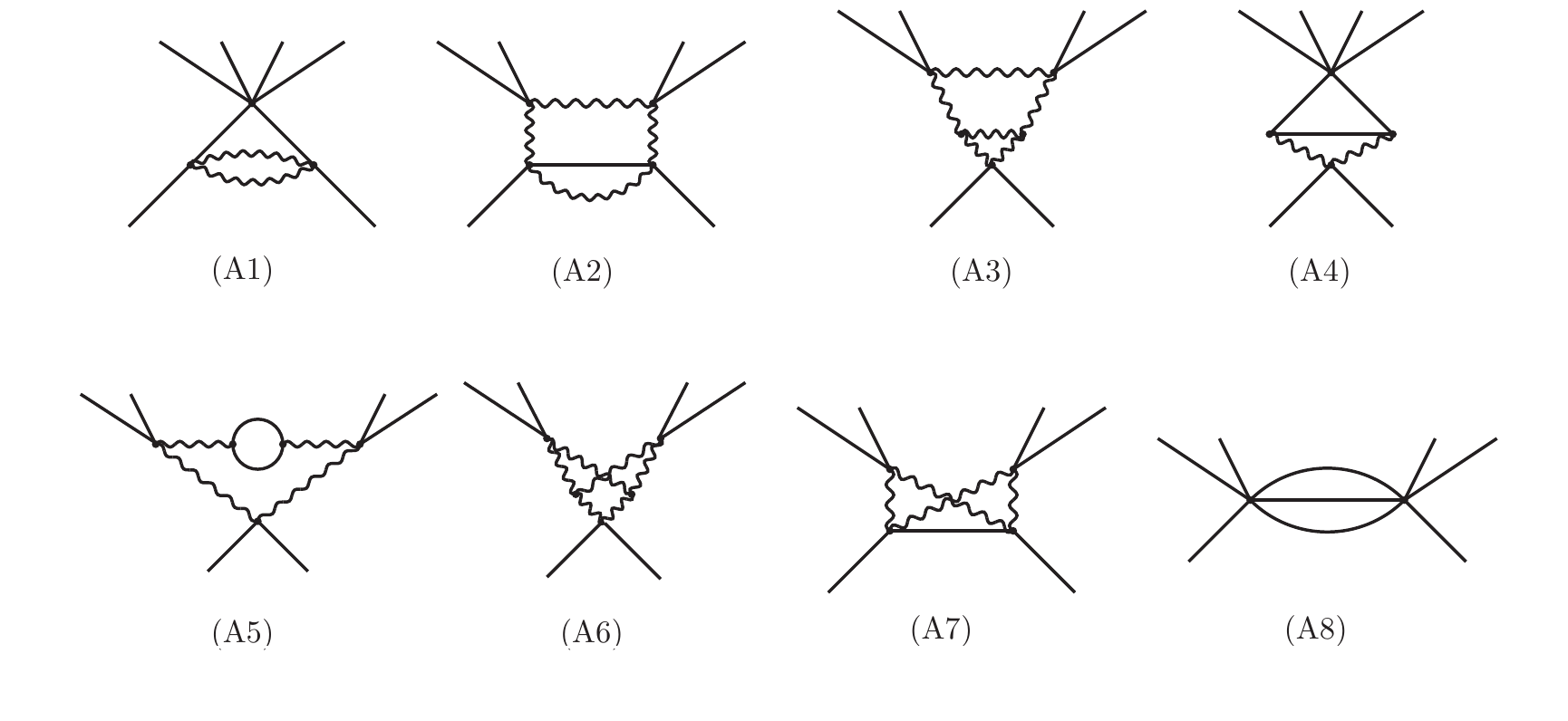}
\end{center}
\vspace{-0.1in}
\caption{\small The diagrams that contribute to the renormalization of the six-point coupling at two loops.}
\label{fig: 6ptdiagrams}
\end{figure}

\begin{figure}
\begin{center}
\includegraphics[width=\textwidth]{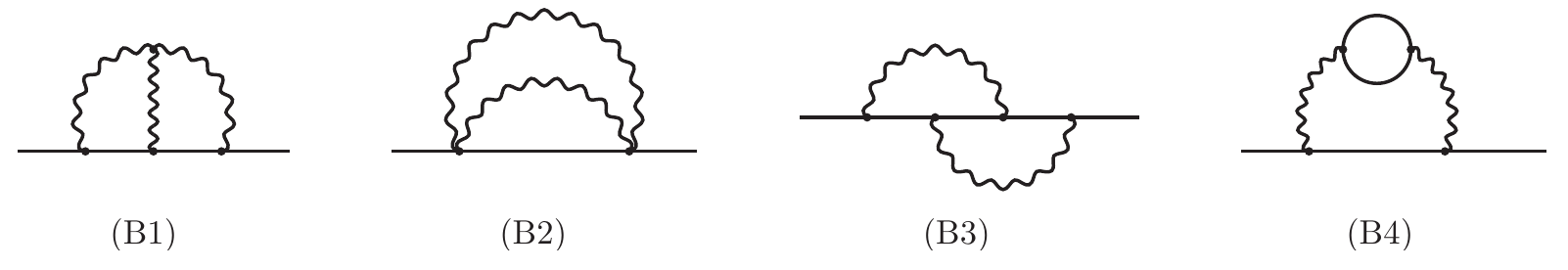}
\end{center}
\vspace{-0.3in}
\caption{\small \vspace{-5in}The diagrams that contribute to the wavefunction renormalization at two loops.}
\label{fig: 2ptdiagrams}
\end{figure}

In this subsection, we study the perturbative RG running of the sextic coupling
in four models, all with bosonic matter: $O(N)$, $U(N)$, $Sp(N,\mathbb{R})$ with ghost matter, and $U(N)$ with ghost matter. The first such computations were performed in \cite{Chen:1992ee}, and the large-$N$ $O(N)$ case was analyzed in \cite{Aharony:2011jz}. In this subsection and the related appendices we use $Sp(N,\R)$ with $N$ even, as opposed to writing $Sp(2N, \R)$, 
so that we will obtain results that are directly comparable to the ones in \cite{Aharony:2011jz}.  We use the Landau gauge $\partial_\mu A^\mu = 0$ and compute to two loops (the first nontrivial order in our dimensional regularization scheme). The relevant diagrams, which are the same as in \cite{Aharony:2011jz}, are shown in Fig.~\ref{fig: 6ptdiagrams} and Fig.~\ref{fig: 2ptdiagrams}. The diagrams for $U(N)$ will be decorated with appropriate arrows.


Our conventions for various terms in the Lagrangian, which are the same as in \cite{Aharony:2011jz}, and the values of all the Feynman diagrams are presented in Appendix \ref{app:diagrams}. Our calculations indicate that \emph{each individual diagram continues $N \rightarrow -N$ from the non-ghost theories}. We emphasize that (at least in Landau gauge) this is not a trivial statement having to do solely with matter loops. There is an intricate structure of minus signs that comes from the anticommuting nature of the fields and (in the case of $O(N)/Sp(N,\mathbb{R})$) from the group theory factors being different. Moreover, the gauge fields themselves are still commuting, and we do not pick up minus signs for different Wick contractions of the gauge field. For example, diagram (A3) has no matter loops yet continues $N\rightarrow -N$ even for the $U(N)$ case.

For $O(N)$ and $Sp(N, \R)$, we find the $\beta$-function\footnote{Note that our result for the $O(N)$ case differs from eq.~(9) in \cite{Aharony:2011jz} because of a slight error in their result for the (A7) diagram. We thank Guy Gur-Ari for confirming this.}
\begin{align}
\beta_{\lambda_6} = \frac{1}{16\pi^2 N^2} \left(
12 \lambda^4 (\pm N-1) - 20 \lambda^2 \lambda_6 (\pm N-1) + \lambda_6^2(\pm3N+22)
\right),
\end{align}
where the upper sign pertains to $O(N)$ with commuting bosons and the lower sign to $Sp(N,\mathbb{R})$ with anticommuting bosons.
For the two $U(N)$ theories, we find
\begin{align}
\beta_{\lambda_6} = \frac{1}{16\pi^2N^2}\left(
3\lambda^4 (\pm 11N+53) -40\lambda^2 \lambda_6(\pm N+1) + 2\lambda_6^2(\pm 3N+11)
\right),
\end{align}
where the upper sign pertains to $U(N)$ with
commuting bosons and the lower sign to $U(N)$ with anticommuting bosons.

We find that the $\beta$-function vanishes identically to leading order in $1/N$. This means that, in the strict $N = \infty$ limit, both couplings $\lambda, \lambda_6$ are exactly marginal for all these models (as in \cite{Aharony:2011jz}, the Chern-Simons coupling $k$ is quantized and expected not to run
). This is due to a nontrivial cancellation between $\O(N^2)$ diagrams. In addition, the results in these two pairs of theories are related by the advertised $N \rightarrow -N$ map. We propose that the fixed lines indexed by $\lambda$ at finite $N$ are dual to the family of de Sitter bulk theories indexed by a parity-violating phase.

Notice that even with $\lambda=0$, the commuting and anticommuting theories show markedly different behavior: the triple-trace interaction is marginally irrelevant for the commuting theories, while it is marginally relevant for the anticommuting theories.


\subsection{Non-perturbative approach in light-cone gauge} \label{sec LC R3}

There exist powerful techniques to evaluate path integrals in a special gauge called the ``light-cone" gauge \cite{Giombi:2011kc, Jain:2012qi}. The Euclidean light-cone coordinates are $x^\pm = (x^1 \pm i x^2)/\sqrt{2}$ and $x^3 = x^3$. Similarly, $p_\pm = (p^1 \mp i p^2)/\sqrt{2}$ and $p^3 = p^3$. We also define $p_s^2 \equiv p_1^2 + p_2^2 =  2 p_+ p_-$ for future convenience.

The Euclidean light-cone gauge is $A_- = (A_1 + i A_2)/\sqrt{2} = 0$.\footnote{The light-cone gauge is formally defined using analytic continuation. This subtlety is treated with great care in, for instance, \cite{Jain:2012qi}, where it is shown that na\"ively setting the above linear combination to zero works, as long as this is done \emph{after} all needed complex conjugations are performed on the gauge field. In other words, the na\"ive extension of the gauge field into the complex plane works as long as it is still treated as an anti-Hermitian matrix in color space.} In this gauge, ghosts decouple and the cubic gauge interaction term vanishes. The action becomes quadratic in $A_3^a$ and linear in $A_+^a$. The latter can consequently be integrated out as a Lagrange multiplier, which fixes the former as a function of the matter fields. This leaves us with an effective potential for a matter-only vector model. As shown in Appendix \ref{Appendix LCgauge}, the partition function for the $U(N)$ CS-ghost boson model \eqref{def ZB} can thus be rewritten as
\algnl{\notag
  Z_B
  &= \int[\d\chi] \exp\Biggr\{-\int_p \bar\chi_i(-p) \chi_i(p) \left[p_s^2 + p_3^2 + \sigma\right] - 2 \pi i \lambda N \int_{P,\,p,\,q}\hspace{-0.8em} \chi^2(P, p) \chi^2(-P, q) C_1(P, p, q)\\
  &\hspace{-2.25em}- (2\pi \lambda)^2 N\hspace{-2em} \int\displaylimits_{P,\,Q,\,R,\,p,\,q,\,r} \hspace{-1.75em} (2\pi)^3\delta^3(P + Q + R)\, \chi^2(P, p)\, \chi^2(Q, q)\, \chi^2(R, r)  \left( C_2(P, Q, R, p, q, r) + \frac{\lambda^b_6}{24\pi^2 \lambda^2}\right)\Biggr\},
}
with the CS level expressed through the 't Hooft coupling $\lambda \equiv N/k$, and with:
\algnl{
  \chi^2(P, p) &\equiv  \frac1N \bar\chi_i \left(\frac P2 - p\right) \chi_i \left(\frac P2 + p\right),\\
  C_1(P, p, q) &\equiv \frac{(P + p + q)_- (-P + p + q)_3}{(p - q)_-},\\
  C_2(P, Q, R, p, q, r) &\equiv \frac{(P - R + 2p + 2r)_- (R - Q + 2q + 2r)_-}{(P + R + 2p - 2r)_- (R + Q + 2r - 2q)_-}~.
}
The partition function has the schematic form
\bel{\label{ZB schemi}
  Z_B \sim \int[\d\chi] \exp\left\{-\bar \chi_i (-\del^2) \chi_i - i\frac\lambda N (\bar\chi_i \chi_i)^2 - \frac1{N^2} (\lambda^2+\lambda_6) (\bar\chi_i \chi_i)^3\right\}.
}
The only difference thus far from the commuting case is the sign of the $(\bar\chi_i \chi_i)^2$ term. At this stage we perform a Hubbard-Stratonovich transformation by inserting
\bel{\label{hs}
  1 = \int [\d\gamma\, \d\mu] \exp\left\{-i\int_{p_1,\, p_2} \mu(p_1,\, p_2)\left[\gamma(p_1,\, p_2) - \bar \chi_i(p_1) \chi_i(p_2)\right]\right\}
}
into the partition function \eqref{ZB schemi}. This leaves an effective action that is quadratic in the $\chi$-fields. After integrating out the $\chi$-fields, the schematic form of $Z_B$  becomes
\bel{\label{ZB schem}
  Z_B \sim \int [\d\gamma\, \d\mu] \exp\left\{N \log\det(-\del^2 - i\mu) - i \frac\lambda N \gamma^2 - \frac1{N^2}( \lambda^2+\lambda_6) \gamma^3 - i \mu \gamma\right\}.
}
The partition function $\~Z_B$ of the commuting boson theory can be written as
\bel{
  \~Z_B\sim \int [\d\gamma\, \d\mu] \exp\left\{-N \log\det(-\del^2 - i\mu) + \frac i{N}\lambda \gamma^2 - \frac1{N^2}( \lambda^2+\lambda_6) \gamma^3 - i \mu \gamma\right\}.
}
The only differences are the signs of the $\log\det$ and $\gamma^2$ terms. This change of signs can be accounted for by sending $N \rightarrow - N$ (while keeping $\lambda$ and $\lambda_6$ fixed)  in all explicit appearances of $N$ in the final expression for the partition function above.

It now remains to show that the dominant large-$N$ saddle-point values of the CS-matter and CS-ghost partition functions map into each other by $N \rightarrow -N$ with $\lambda$ and $\lambda_6$ fixed.  That this is the case follows from the analysis of \cite{Aharony:2012nh}, which shows that there exists only one saddle point (solution to the gap equations) at large $N$. Corrections in powers of $1/N$ are then expected to simply transform via $N \rightarrow - N$. Note that at finite $N$ we may also expand in powers of $\lambda$ to get the same $N \rightarrow - N$ relation between ordinary and ghost matter theories, term-by-term in $\lambda$.

So far we have shown that the partition function of the $U(N)$ ghost boson model on $\R^3$ is related to the partition function of the $U(N)$ ordinary boson model by $N \rightarrow -N$. We must do the same for all correlators, which requires adding sources. Consider the generating functional
\bel{
  Z_B[J_\mu^a, J_i] = \int [\d A\, \d\chi] \exp\left\{-S\_{CS} - S\_{B} -  N \int\d^3 x\, J_\mu^a A^a_\mu - \int \d^3 x (J_i \chi_i +  \bar J_i \bar\chi_i)\right\}.
}
The explicit factor of $N$ in front of the $J_\mu^a A_\mu^a$ term is inserted for later convenience. We may consider this object after fixing the gauge $A_-^a = 0$; then the only two relevant components of $J_\mu^a$ are $J_-^a$ and $J_3^a$. Any correlator can be expressed as a series of differential operators $-\frac 1N \fder{}{J_\mu^a}$ and $-\fder{}{J_i}$ acting on $Z_B[J_\mu^a, J_i]$ and then setting all sources to zero. (In what follows we do not discuss contact terms.) With these sources turned on, we may now again integrate out the gauge fields, getting a functional that can be schematically written as
\algnl{\notag
  Z_B[J_\mu^a, J_i] \sim \int [\d\chi] \exp\Bigg\{-S\_{old} & -  \lambda J_- \left(\bar\chi_i \chi_i - i\frac\lambda N(\bar\chi_i\chi_i )^2\right) - \\
  & \qquad - \lambda^2 J_-^2 (\bar\chi_i\chi_i) -  \lambda J_3 (\bar\chi_i\chi_i) - (J_i\chi_i +  \bar J_i \bar\chi_i) \Bigg\}.
}
Note that we have dropped all prefactors and momenta; for instance, we have written terms like $\bar\chi_i T_{ij}^a J_-^a \del_+ \chi_j$ as $ J_- \bar\chi_i\chi_i$ and $J_-^a \bar \chi_i T^b_{ij} \chi_j \bar\chi_k T^{a}_{kl}T^b_{lm}\chi_m$ as $J_- (\bar\chi_i\chi_i)^2$, with Lie algebra indices summed over, when possible, using the Fierz identity \eqref{UNFierz}. The action $S\_{old}$ is the same one found above in eq.~\eqref{ZB schemi}. The only difference of the additional terms, compared to the ordinary boson case, is the sign of the $J_- (\bar\chi_i\chi_i)^2$ term, whose coefficient is proportional to $1/N$. This sign comes from the difference in statistics.

We can now perform a Hubbard-Stratonovich transformation analogous to the case with no sources by inserting \eqref{hs} into the path integral. An additional complication enters because the functional determinant with sources turned on is no longer a determinant of the diagonal matrix $\delta_{ij}$ in color space. However, the $\log\det$ can be expanded perturbatively in the infinitesimal sources, giving
\bel{
\label{eq-zbsource}
  Z_B[J_\mu^a, J_i] \sim \int[\d\gamma\, \d\mu] \exp\left\{W\_{old}[\gamma, \mu] + J^a \Tr T^a + J^a J^b \left[\Tr(T^a T^b) + \Tr T^a \Tr T^b\right] + \ldots \right\}.
}
As before, we have dropped all prefactors and momentum integrals (for instance, the coefficients of the $J^a$ terms also depend on $\mu$ and $\gamma$). We use $W\_{old}$ to denote the exponent appearing in the generating functional \eqref{ZB schem} of the no-source theory, with $Z_B[0] = \int [\d\gamma \, \d \mu] e^{W\_{old}[\gamma,\, \mu]}$. The generating functional of the corresponding theory with ordinary matter is
\bel{
\label{eq-zbtildesource}
  \~Z_B[J_\mu^a, J_i] \sim \int[\d\gamma\, \d\mu] \exp\left\{\~W\_{old}[\gamma, \mu]  - J^a \Tr T^a - J^a J^b \left[\Tr(T^a T^b) + \Tr T^a \Tr T^b\right] - \ldots \right\},
}
with $\~W\_{old}$ being obtained from $W\_{old}$ by $N \rar - N$, as described above. All coefficients dropped in this expression for $\~Z_B[J_\mu^a, J_i]$ are
given by applying the mapping $N \rightarrow -N$  to the coefficients in the expression for $Z_B[J_\mu^a, J_i]$.

By comparing (\ref{eq-zbsource}) and (\ref{eq-zbtildesource}),  we see that all terms containing sources of gauge fields change the overall sign as the statistics is changed. This is due to the change of sign of the log det term that arises by integrating out matter, but there is no explicit factor of $N$ to soak up this sign. However,
any correlator will contain a number of derivatives of the form $\frac1N \fder{}{J_\mu^a}$ acting on $Z_B[J_\mu^a, J_i]$. This means that every differentiation with respect to $J^a_\mu$ will pick up exactly one statistics-dependent sign. Since each such differentiation comes with a factor of $1/N$, this is exactly what should happen if the correlators (modulo contact terms) are to be related by $N \rar -N$ upon a change of matter statistics.

Completely equivalent considerations can be applied to $U(N)$ ghost fermion models, and the same results as above are found. We already know that ordinary bosonic and fermionic CS models can be bosonized into each other on $\R^3$ (at least at large $N$) \cite{Aharony:2012nh}, and hence we conclude that the ghost CSM models follow suit by applying the $N\rightarrow -N$ map.

Finally, we note that, using the same techniques as above, the $O(2N)$ regular boson model can be shown to map via $N \rightarrow - N$ to the $Sp(2 N,\mathbb{R})$ ghost boson model. The fermions obey this map as well, leading to bosonization of the $Sp(2N,\mathbb{R})$ ghost theories. 

The dualities discussed in this section fit into a bigger web of dualities. These are all shown on Fig.~\ref{fig arrowsdiag}.

\subsection{$\Psi\_{HH}$ at finite $\lambda$?}

We have argued that the correlators of gauge-invariant operators on $\R^3$ in the theory with ghost vector matter are related by an analytic continuation $N \rightarrow -N$, with $\lambda$ and $\lambda_6$ fixed, to the cases with ordinary vector matter. According to the dS/CFT correspondence, these theories are dual to parity violating higher-spin bulk de Sitter theories. Their partition function, as a function of the sources, is computing the wavefunctional $\Psi\_{HH}$ in the Hartle-Hawking state, as a function of the late-time profiles of the bulk fields. When the theory resides on {$\mathbb{R}^3$} and we take large $N$, there is a bosonization analogous to the one discussed in \cite{Aharony:2012nh} relating theories with anticommuting free (critical) bosons to theories with commuting critical (free) fermions.

In \cite{Anninos:2012ft} it was observed that $Z\_{CFT}[\sigma]$ grew for large values of the uniform source $\sigma$ of the {$\bar{\chi}_i\chi_i $} operator, for the free anticommuting bosonic theory on $S^3$ (with $\lambda =\lambda_6= 0$). This implies a growth of the wavefunction for large values of the bulk scalar. (More precisely, it is a function of the fast-falling profile $\beta$ of the late-time expansion $\varphi \sim \alpha \eta + \beta \eta^2$ of the bulk scalar $\varphi$. The critical anticommuting scalar computes $\Psi\_{HH}$ for the slow-falling profile $\alpha$ of the bulk scalar.) Does this phenomenon persist at finite $\lambda$? From the point of view of bosonization, one might expect that the $\lambda \to 1$ case can be obtained by considering the partition function of free commuting spinors at $\lambda \to 0$. (For this to be true one would have to establish that bosonization holds on $S^3$ also, which is a natural speculation.) An intriguing possibility is that only a discrete set of $\lambda$'s lead to normalizeable wavefunctions. We leave the exploration of $Z\_{CFT}$ on $S^3$ at finite $\lambda$, with various sources turned on, to future work.

\section{CS-ghost dualities on $S^1 \times S^2$}\label{sec Thigh}

In this section we study the non-perturbative aspects of the CS-ghost models with $U(N)$ symmetry by computing their ``thermal" partition functions on $S^1 \times S^2$. We assume that the volume of the spatial manifold is large, so our results are applicable to $S^1 \times \Sigma_g$ for any genus-$g$ Riemann surface $\Sigma_g$ \cite{Aharony:2012ns}. We will henceforth focus exclusively on the $U(N)$ models. From the holographic point of view we are computing the Hartle-Hawking wavefunction as a function of an $S^1 \times S^2$ boundary with all other bulk fields turned off at late times. We will refer to $\beta$, the size of the $S^1$, as an inverse temperature in accord with the literature, but we emphasize here that this is simply shorthand and there is no thermodynamic interpretation of the $S^1$ factor.

\subsection{Preliminary remarks}

The first computation of a thermal partition function for ghost theories was done for the free ($\lambda = \lambda^b_6 = 0$) theory of $U(N)$ anticommuting scalars \cite{Anninos:2012ft}. Let us focus on this special case. The matter in this theory does not interact with gauge fields but does couple to the holonomy of the gauge field along the thermal $S^1$. Integrating out this holonomy at low temperatures imposes the singlet constraint on the matter operators, while at high temperatures the holonomy eigenvalues undergo a Gross-Witten-Wadia transition and eventually clump at a single point, thereby lifting the singlet constraint, as is expected at high temperatures \cite{Shenker:2011zf}.

From an alternative point of view, integrating out the matter gives an effective potential for this holonomy. At high temperatures, $T \gtrsim \sqrt N$, the holonomy is governed just by this effective potential, and the partition function takes the form
\bel{\label{Z DDD}
  Z = \int [\d \alpha_i] e^{-4 T^2 \sum_i \sum_{m = 1}^\infty \frac{(-1)^m}{m^3}\cos m \alpha_i },
}
where $\alpha_i$ are the $N$ eigenvalues of the holonomy, and the spatial $S^2$ is chosen to be of unit radius, so its volume is $V_2 = 4\pi$. Both minus signs in the exponent come from the anticommuting nature of the matter fields; the high-temperature effective potential for the commuting matter theory is retrieved by flipping both minuses to pluses \cite{Shenker:2011zf}. It is important to keep track of the origin of these signs: the overall minus sign is there because the path integral over matter gives a positive power of the one-loop determinant, and the $(-1)^m$ factors are there because the matter has antiperiodic boundary conditions along the thermal circle. To see the latter point more clearly, note that $(-1)^m \cos m\alpha = \cos m(\alpha + \pi)$; the phase shift by $\pi$ along the thermal circle precisely accounts for the antiperiodic boundary conditions.

At large $N$ and $T$, the eigenvalues $\alpha_i$ all clump together to minimize the free energy. For the theory \eqref{Z DDD}, they clump around $\alpha = 0$, where the saddle-point value of the partition function is
\bel{\label{F DDD}
  Z = e^{-4N T^2 \sum_{m = 1}^\infty \frac{(-1)^m}{m^3}} = e^{3N T^2 \zeta(3)}.
}
This is to be compared to the commuting matter result $Z = e^{4N T^2 \zeta(3)}$, which is \emph{not} related to the anticommuting result by $N \rightarrow -N$, as pointed out in \cite{Anninos:2012ft}.

Even though the $N \rightarrow -N$ map does not work, bosonization between anticommuting scalars and commuting fermions could still hold. Studying this requires a more involved analysis at finite $\lambda$, and at present we only have the technology to study the $T \gtrsim \sqrt N$ regime, in which quantum fluctuations of the holonomy may be ignored. The development of the tools necessary for this high $T$ computation was completed in \cite{Aharony:2012ns} with the discovery of the non-trivial $\lambda > 0$ effect (``Pauli exclusion'') that forbids holonomy eigenvalues from clumping together \emph{too} tightly around a single clumping point, forcing them instead to be uniformly distributed over an interval of length $2\pi|\lambda|$.\footnote{We assume that the temperature is high enough so that (using the nomenclature of \cite{Jain:2013py}) we do not encounter other gaps in the eigenvalue density.} We closely follow and extend the analysis of \cite{Aharony:2012ns} to dynamical (varying) clumping points of eigenvalues; this is a natural and (as it turns out) necessary move, if our goal is to study the wider perspective of the saddle point structure of CS-matter theories. We will demonstrate that saddle points may occur for clumpings at either $0$ or $\pi$ (or both), and that bosonization dualizes only \emph{some} of the saddles found in bosonic and fermionic theories.

Concretely, we tackle CS-ghost theories at high temperatures $T = 1/\beta$ using tools similar to those that proved effective in Section \ref{sec LC R3}. We fix the light-cone gauge, integrate out all the gauge fields except the holonomy (whose eigenvalues are uniformly distributed around a dynamical clumping point), obtain an effective action for a vector model in a background field, and then solve the saddle-point equations. This way, studying the entire CS-ghost theory is reduced to minimizing the free energy as a function of the clumping position, the thermal mass, the Legendre transform parameter (present if we are studying the critical theory), and certain discrete variables; see Appendix \ref{Appendix HighT}.

Before we proceed, a word of caution about the saddle-point analysis is in order. We do not perform the full steepest descent analysis. This is a very subtle subject in these theories, as it (for example) involves studying the interplay between renormalization and asymptotic behavior in a space of several complex parameters which figure in the Morse function. Instead, we content ourselves with finding all saddle points and choosing the single dominant one. As a consistency condition, we note that this dominant saddle is always the one that is continuously connected to the free matter theory by dialing $\lambda$ to 0. At the free field point, we may obtain the partition function by treating the CFT as a high-temperature gas of particles, and this computation agrees with the partition function at the dominant fixed point. As $\lambda$ is increased, no other saddle point takes over in dominance, and therefore no Stokes phenomena should be encountered.\footnote{We thank Steve Shenker for many discussions of this issue.} We will therefore assume that the only physical saddle point is the dominant one. This issue certainly deserves further study along the lines first promoted in \cite{Witten:2010cx, Harlow:2011ny}, and we hope to return to it in future work.

\subsection{General properties of saddle points}

After the dust from Appendix \ref{Appendix HighT} settles, the anticommuting bosonic theory is found to be governed by an ``off-shell'' free energy
\bel{\label{FB}
  F_{B\pm}\left(\sigma, \mu_B^2, \avg{a}\right) = \frac {NV_2T^2}{2\pi} \Bigg\{\frac{\mu_B^3}3 \pm \frac23 \frac{\left(\mu^2_B - \sigma \beta^2\right)^{3/2}}{\sqrt{\lambda^2 + \lambda_6^b/8\pi^2}} + \frac2{\pi \lambda} \int_{\mu_B}^\infty \d y\, y\,\trm{Im}\, \trm{Li}_2\left(-e^{-y + i\pi\lambda + i \avg{a}} \right)\Bigg\}
}
with $\avg a \in \{0,\pi\}$, $\mu_B \geq 0$, and $\trm{Li}_n(x) \equiv \sum_{m = 1}^\infty \frac{x^m}{m^n}$ is the $n$-th polylogarithm. The free energy is put ``on shell'' by extremizing it w.r.t.~all of its explicitly written arguments, including the $\pm$ sign. (This sign comes from an intermediate step in solving the Schwinger-Dyson (gap) equations, see eq.~\eqref{origin of pm}.) The saddle-point values of $F$ and $\avg a$ will be indicated by ${}\^{reg}$ or ${}\^{crit}$ superscripts denoting the theory in question.

If we include clumping point dynamics in the ordinary commuting boson model of \cite{Aharony:2012ns}, the off-shell free energy is
\bel{
  \~F_{B\pm}\left(\sigma, \mu_B^2, \avg{a}\right) = - \frac {NV_2T^2}{2\pi} \Bigg\{\frac{\mu_B^3}3 \pm \frac23 \frac{\left(\mu^2_B - \sigma \beta^2\right)^{3/2}}{\sqrt{\lambda^2 + \lambda_6^b/8\pi^2}} + \frac2{\pi \lambda} \int_{\mu_B}^\infty \d y\, y\,\trm{Im}\, \trm{Li}_2\left(e^{-y + i\pi\lambda + i \avg{a}} \right)\Bigg\}.
}
The commuting boson with $\avg a = 0$ has the same thermal mass and the same absolute free energy as the anticommuting boson with $\avg a = \pi$, and vice versa; the only difference lies in the signs of the self-energies. An immediate consequence of this is that the same points in the parameter space of $\mu_B^2$, $\sigma$, and the $\pm$ sign are saddles in both the commuting and anticommuting case. The signs of the free energies at these saddles change as the statistics is changed, though, so the ghost theories may (and generally do) have different dominant saddle points.

The fermionic off-shell free energy is, according to Appendix \ref{Appendix HighT},
\algnl{\notag
  F_{F\pm}\left(\sigma, \mu_F^2, \avg{a}\right)
  = -\frac{N V_2 T^2}{2\pi} \Biggr\{\frac{\mu_F^3}3 &\frac{\lambda \pm 1}{\lambda} + \frac{\sigma \beta \mu_F^2}{2\lambda} - \frac{(\sigma\beta)^3}{6\lambda} + \frac{\pi \lambda_6^f(\sigma \beta)^3}{3} + \\ \label{FF}
  &\qquad\qquad\qquad + \frac2{\pi \lambda} \int_{\mu_F}^\infty \d y\, y \,\trm{Im\,Li_2}\!\left(e^{-y + i\pi \lambda + i \avg a} \right) \Biggr\}.
}
Again, we have $\avg a \in \{0,\pi\}$ and $\mu_F \geq 0$. The upshot is the same as in the bosonic case: the off-shell free energies of commuting and anticommuting fermionic theories differ only by an overall sign, and both theories have saddle points at the same values of $\mu_F$, $\sigma$, and the $\pm$ sign.

In the following two subsections we collect information about all saddle points in all theories of interest at arbitrary values of couplings $\lambda$ and $\lambda_6$ (as applicable). We will show that bosonization fails to work in a very specific way: all bosonic saddle points will be mappable to fermionic saddle points, but there exist fermionic saddle points which are not dualized to anything in the bosonic theory. In particular, the dominant fermionic saddle has no bosonic dual, in contrast to the previously studied cases.

As already noticed in \cite{Aharony:2012ns}, it is especially tidy to express all results in terms of the function
\bel{
  \F(\mu, z) = \frac1{\pi^2} \trm{Im} \left[\frac{\mu^2}3 \trm{Li}_2\left(z e^{-\mu}\right) + \int_{\mu}^\infty \d y\,y\,\trm{Li}_2\left(z e^{-y}\right) \right].
}
In all cases of note, $z$ will be a pure phase of the form $\pm e^{i\pi \lambda}$, and the sign of $\F(\mu, z)$ will equal the sign of the imaginary part of $z$.

\subsection{Critical boson and regular fermion}

\begin{figure}
  \centering
  \includegraphics[width=0.45\textwidth]{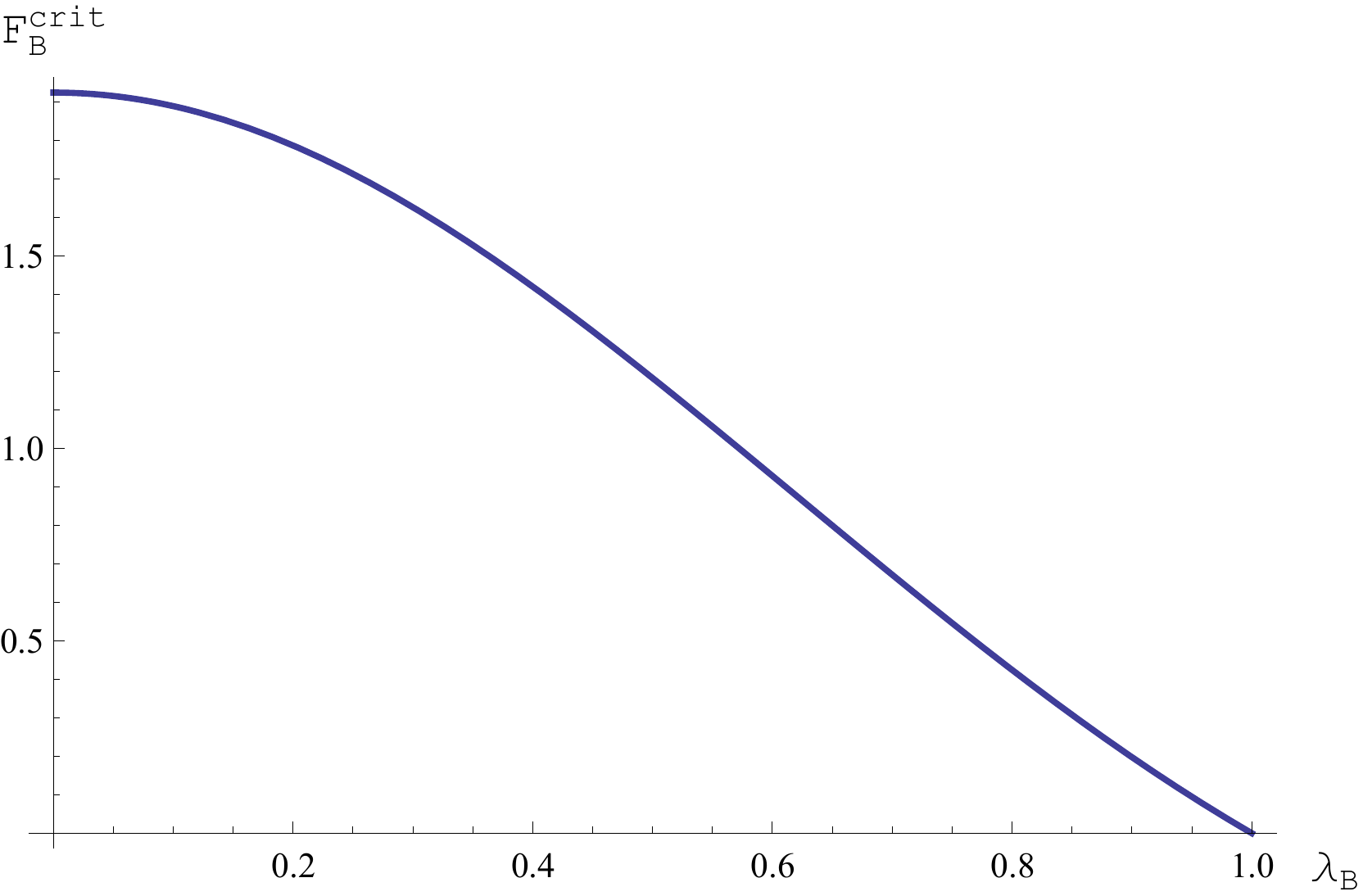}
  \hspace{0.02\textwidth}
  \includegraphics[width=0.45\textwidth]{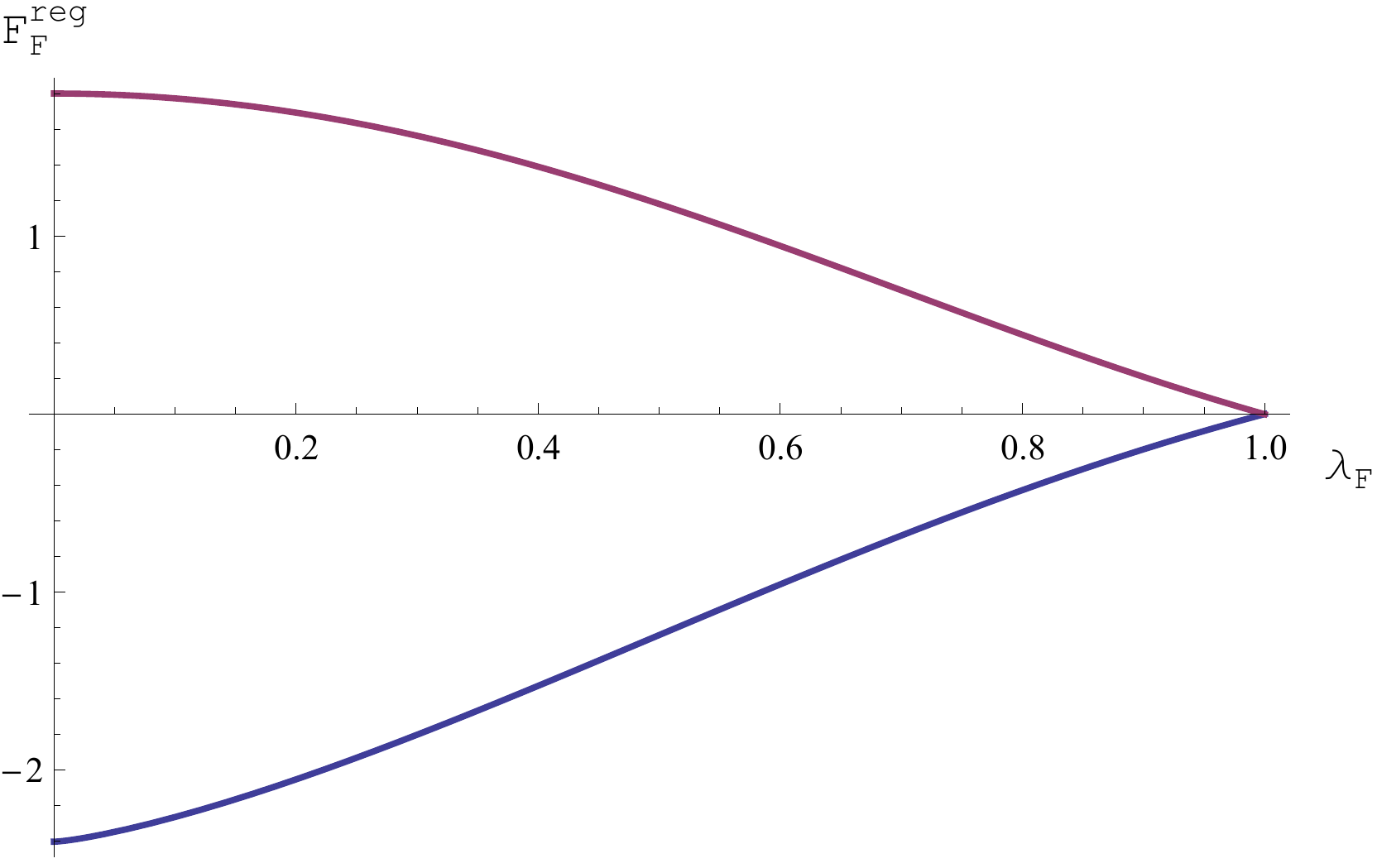}\\
  \caption{\small Free energies of the critical boson (left) and regular fermion (right) at all applicable saddle points, plotted with arbitrary normalization as functions of $\lambda_B$ (left) and $\lambda_F$ (right). On the right (regular fermion) plot, the blue/bottom-most line corresponds to the ``$+$'' saddle which is dominant in this $\lambda_F \geq 0$ regime, and the purple/top-most line corresponds to the ``$-$'' saddle, which is sub-dominant.}\label{fig FBCritFFReg}
\end{figure}

The simplest theory is the critical anticommuting boson with $N_B$ particles and at 't Hooft coupling $\lambda_B$. Its free energy is given by setting $\sigma = \mu^2_B T^2$ in \eqref{FB} and solving the gap equation. This means that the term containing $\lambda_6^b$ drops out of the free energy, confirming that the six-point operator is not a marginal perturbation of the Wilson-Fisher theory. The saddle-point equation for $\mu_B$ is
\bel{
  \frac1{\mu_B}\pder{F\^{crit}_B}{\mu_B} = \mu_B - \frac{2}{\pi\lambda_B} \trm{Im\,Li}_2\left(-e^{-\mu_B + i\pi \lambda_B + i\avg a} \right) = 0.
}
This equation has no solutions for $\avg a = 0$, as the center term is always positive in that case, so we conclude that $\avg a_B\^{crit} = \pi$. This implies that the free energy of the critical anticommuting boson is precisely the negative free energy of the critical commuting boson. This theory, thus, has only one saddle point. The on-shell free energy can be written as:
\bel{
  F_B\^{crit} = \frac{N_B V_2T^2}{\lambda_B} \F\left(\mu_B, e^{i\pi \lambda_B}\right),
}
with
\bel{
  \mu_B = \frac2{\pi\lambda_B} \trm{Im\,Li}_2\left(e^{- \mu_B + i \pi \lambda_B}\right).
}
The free energy is plotted on Fig.~\ref{fig FBCritFFReg}. Note that the free energy goes to zero at $\lambda_B \rar 1$, just like in the ordinary commuting case.

The natural dual to the critical anticommuting boson is the regular commuting fermion. Its free energy is obtained by setting $\sigma = 0$ in \eqref{FF} and it has two saddle points, one for each choice of sign in the $\mu_F^3(1\pm \lambda_F)/3\lambda_F$ term in the off-shell free energy. (We have commented on this in the previous subsection.) The saddle point values are
\bel{
  F\^{reg}_{F\pm} = -\frac{N_F V_2T^2}{\lambda_F} \F\left(\mu_{F\pm}, \pm \trm{sgn}(\lambda_F) e^{i \pi \lambda_F}\right),
}
with
\bel{
  \mu_{F\pm}\left(1\pm \frac1{\lambda_F}\right) = \frac2{\pi\lambda_F} \trm{Im\,Li}_2\left(\pm\trm{sgn}(\lambda_F)e^{- \mu_{F\pm} + i \pi \lambda_F}\right).
}
The clumping point $\avg a_{F\pm}\^{reg}$ is dialed to let the gap equation have solutions for a given sign $\pm$ in the free energy, and for the two cases above the clumping occurs at $\avg a\^{reg}_{F\, \trm{sgn}(\lambda_F)} = 0$ and at $\avg a\^{reg}_{F\, -\trm{sgn}(\lambda_F)} = \pi$. For a given $\lambda_F$, the sign of the free energy is $\mp \trm{sgn}(\lambda_F)$, and because of this sign difference it is possible to clearly distinguish the dominant from the subdominant saddles.

To simplify matters, let us take $\lambda_F \geq 0$. The free energies at both saddles are plotted on Fig.~\ref{fig FBCritFFReg}. Now the ``$+$'' fermionic saddle is the dominant one. If the critical boson and regular fermion are to be bosonized into each other, they must have $\mu_{F\pm} = \mu_B$. Consistency of gap equations makes this possible only if we pick the lower sign. In this case we have $\lambda_F -1 = \lambda_B$ and $\avg{a}\^{reg}_{F-} = \avg{a}\^{crit}_B = \pi$.\footnote{It is reassuring that the clumping points of eigenvalues, being classical observables, are equal on both sides of the duality.} The free energies are equal if we choose $N_B/\lambda_B = - N_F/\lambda_F$, and since we have chosen $\lambda_F \geq 0$, this means that bosonization can hold if we require
\bel{\label{bosonization}
  \frac{N_B}{|\lambda_B|} = \frac{N_F}{|\lambda_F|}, \quad |\lambda_B| + |\lambda_F| = 1,\quad \trm{sgn}(\lambda_B) = -\trm{sgn}(\lambda_F).
}
It is straightforward to verify that this rule also works for $\lambda_F \leq 0$ (the fermionic theory labeled by + is now dual to the bosonic theory, but we still have $\avg{a}\^{crit}_B = \avg{a}\^{reg}_{F+} = \pi$ and $\mu_B = \mu_{F+}$). These are the \emph{same} bosonization rules that hold in the ``ordinary statistics'' case \cite{Aharony:2012ns}. However, there is a major issue here: the critical boson is dual to a \emph{subdominant} saddle point of the regular fermion theory. In our large-$N$ scheme, we pick the dominant saddle point only, and hence we conclude that bosonization is violated.

Note that this issue did not arise in the ordinary matter theories studied in \cite{Aharony:2012ns}, where the dominant saddles mapped to each other and were attained at $\avg a = 0$ in all cases. It is because of this that bosonization in those theories was demonstrated without taking into account the possible clumping at $\avg a = \pi$. In our case, however, letting the clumping position be dynamical is necessary; for instance, the critical ghost boson has no saddle points at $\avg a = \pi$, and in other theories the dominant saddle is also often the one with $\avg a = \pi$.

\subsection{Regular boson, critical fermion, and a phase transition}

\begin{figure}
  \centering
  \includegraphics[width=0.45\textwidth]{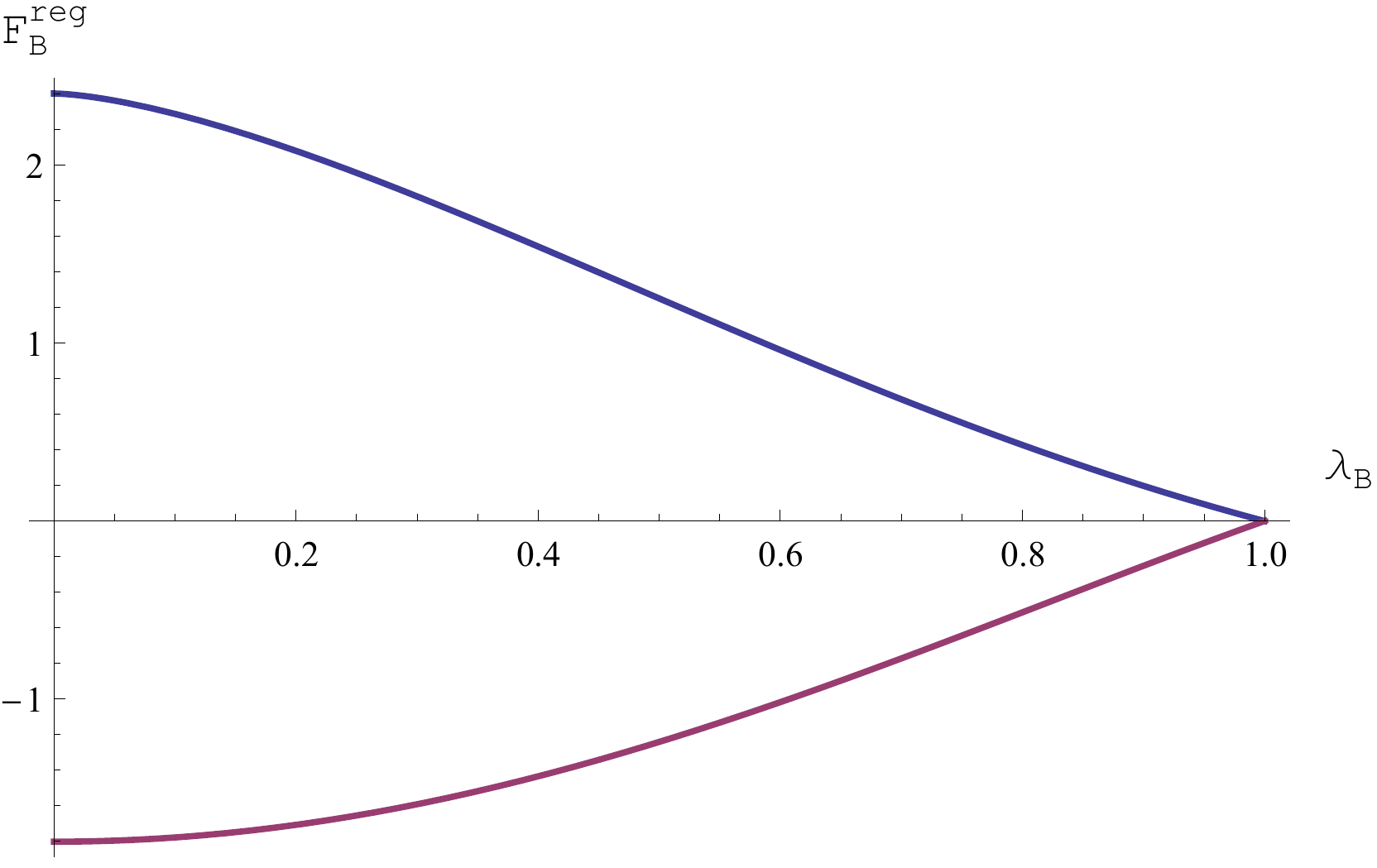}\\
  \caption{\small The free energy of the regular bosonic model, plotted in arbitrary normalization as a function of $\lambda_B$ at $\lambda_6^b = 0$ (i.e.~at $\hat\lambda_B = \lambda_B$). The purple/bottom line corresponds to the dominant, ``$-$'' saddle, and the blue/top line corresponds to the sub-dominant, ``$+$'' saddle.}\label{fig FBReg}
\end{figure}

The analysis of the other pair of theories is more involved, but similar phenomena can be found. The off-shell free energy of the regular boson with a marginal deformation is
\bel{
  F_{B\pm}\^{reg} = \frac{N_B V_2T^2}{\lambda_B} \F\left(\mu_{B\pm}, \trm{sgn}(\hat\lambda_B\pm 2) e^{i\pi \lambda_B}\right),
}
with
\bel{
  \hat \lambda_B \equiv \sqrt{\lambda^2_B + \frac{\lambda_6^b}{8\pi^2}} \geq 0, \qquad \lambda_6^b \geq - 8\pi^2\lambda^2_B,
}
and
\bel{\label{gap Breg}
  \mu_{B\pm}\frac{\hat\lambda_B \pm 2}{\hat\lambda_B} = \frac2{\pi \lambda_B} \trm{Im\,Li}_2\left(\trm{sgn}(\hat\lambda_B\pm 2) e^{- \mu_{B\pm} + i \pi \lambda_B}\right).
}
As before, the clumping position is chosen such that the gap equation has a solution, and this time the choice is $\exp\left(i\avg{a}\^{reg}_{B\pm}\right) = - \trm{sgn}(\hat\lambda_B \pm 2)$. Note that $\avg a\^{reg}_{B+} = \pi$ at any coupling values, while $\avg a\^{reg}_{B-} = \pi$ for $\hat\lambda_B >2$ and $\avg a\^{reg}_{B-} = 0$ for $0 \leq \hat\lambda_B < 2 $. Numerical evaluation shows that $F_{B-} < F_{B+}$ at all couplings, and thus the ``$-$'' saddle is always dominant, with a second-order phase transition at $\hat \lambda = 2$ where the clumping point $\avg a\^{reg}_{B-}$ jumps from $0$ to $\pi$ as $\hat\lambda_B$ is increased. The free energies at both saddles are plotted on Fig.~\ref{fig FBReg}.

At $\hat\lambda_B = 2$, i.e.~on the line
\bel{
  \lambda_6^b(\lambda_B) = 8\pi^2(4 - \lambda_B^2),
}
the gap $\mu_{B-}$ diverges as $-\log|2 - \hat\lambda_B|$, but the free energy stays finite --- in fact, it becomes zero and changes sign as $\hat \lambda$ is dialed across $2$. It is apparent that near this transition, $\pder{\mu_{B-}}{\hat \lambda_B}$ behaves as $|2 - \hat\lambda_B|^{-1}$, leading to a divergence in the second derivative of $F_{B-}$ through a term of the form $\pdder{\F}{\hat\lambda_B} \sim \pdder{\F}{\mu_{B-}} \Big(\pder{\mu_{B-}}{\hat \lambda_B}\Big)^2$.  The transition is therefore second-order. This is a finite-$\lambda$, high-$T$ manifestation of the a phenomenon long ago discovered by Bardeen, Moshe, and Bander in an ungauged $\phi^6$ commuting boson model at zero temperature \cite{Bardeen:1983rv} and very recently in the CS-boson theory at zero temperature \cite{Bardeen:2014paa}. The transition point $\lambda_6^b = 32\pi^2$ in the $\lambda = 0$, $T = 0$, ordinary boson theory was associated to the spontaneous breakdown of scale invariance due to the generation of a mass term for the scalar particle. At high $T$ we find no evidence of conformal symmetry being broken; as we cross the transition point, however, we only preserve the saddle point if we change $\avg a$ to its other allowed value. This maneuver is not allowed at $T = 0$, where there is no thermal circle. At the transition itself, the matter has an infinite gap and the partition function is unity; this means that the matter is screened away and we are left with the pure CS theory.

We thus conclude that this is a second-order phase transition that can be diagnosed by $\avg a$, the eigenvalue clumping point of the Polyakov loop. The transition line parametrized by $\lambda$ is not a line of critical points, as the gap becomes infinite and not zero; it is perhaps better to use the term ``topological points.'' This interesting phenomenon exists in the ordinary matter theory of \cite{Aharony:2012ns} as well, and it would be fascinating to explore it further.

\begin{figure}
  \centering
  \includegraphics[width=0.45\textwidth]{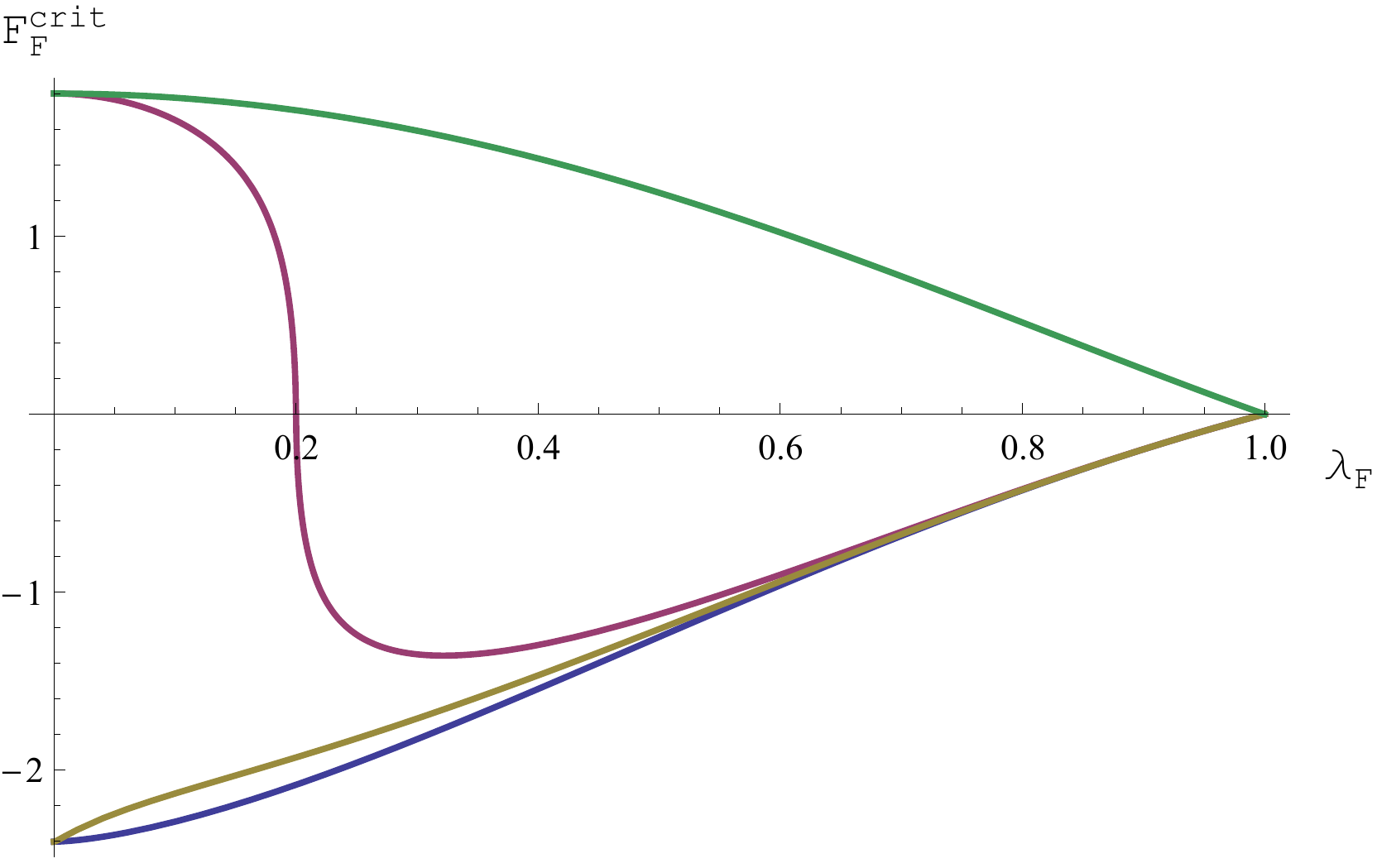}
  \hspace{0.02\textwidth}
  \includegraphics[width=0.45\textwidth]{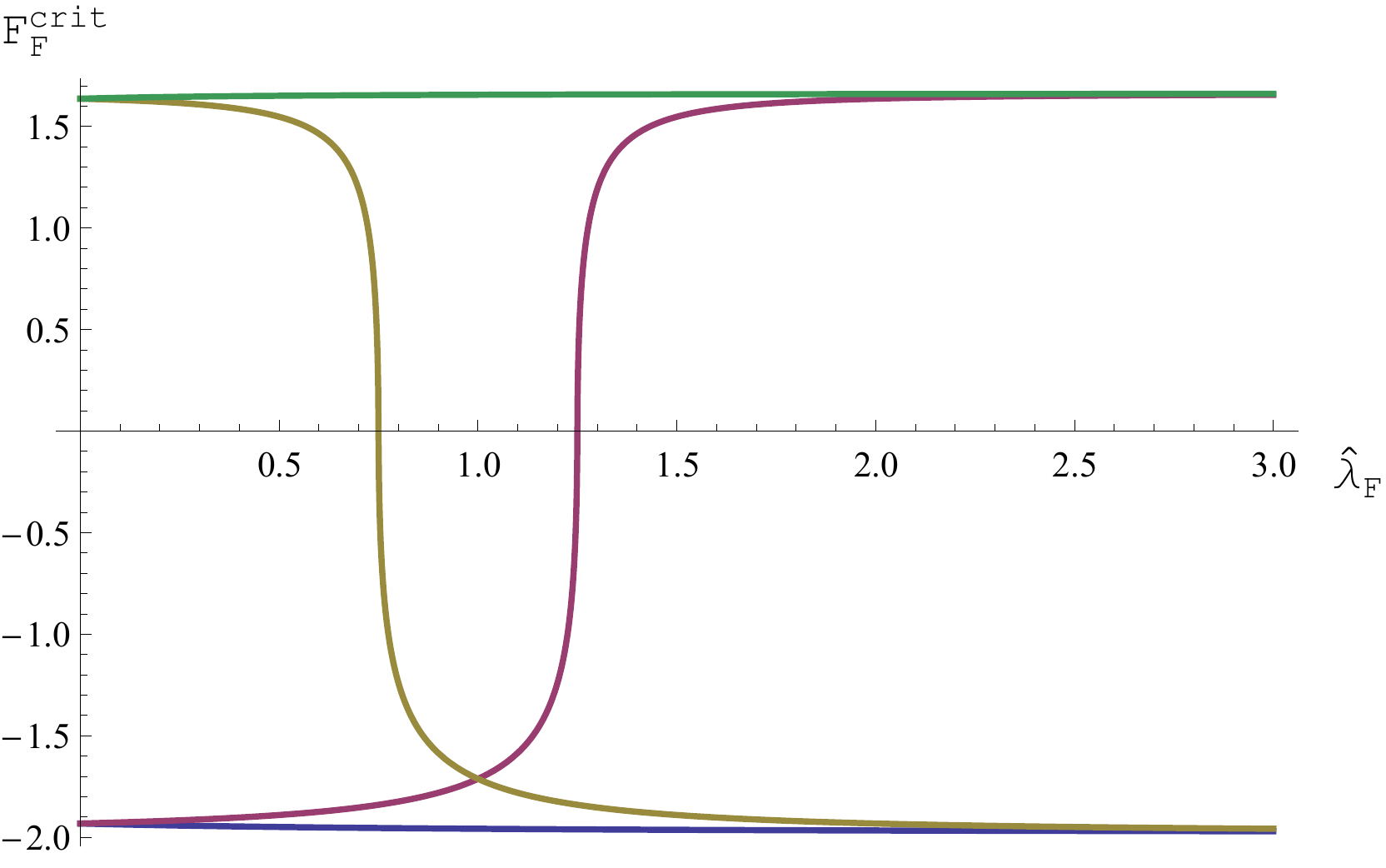}\\
  \caption{\small Free energies of the critical fermion at all four saddle points, plotted with arbitrary normalization as functions of $\lambda_F$ at fixed $\hat\lambda_F = 1.2$ (left) and as functions of $\hat \lambda_F$ at fixed $\lambda_F = 1/4$ (right). The blue/bottom-most line corresponds to the dominant ``$++$'' saddle on both plots. The green/top-most line corresponds to the ``$--$'' saddle, which is the least dominant in the entire parameter space. The purple line (color online) corresponds to the ``$+-$'' saddle, which reaches the phase transition at $\hat\lambda_F = 1 + \lambda_F$, as can be seen on both plots. The beige line corresponds to the ``$-+$'' saddle, which undergoes the phase transition at $\hat \lambda_F = 1 - \lambda_F$. }\label{fig FFCrit}
\end{figure}

The critical fermion theory has an even more intricate structure. Integrating out the auxiliary field $\sigma$ in \eqref{FF} is tantamount to setting $|\sigma|\beta = \mu_F\hat\lambda_F$ with
\bel{
  \hat\lambda_F \equiv \frac1{\sqrt{1- 2\pi \lambda_F \lambda_6^f}} \geq 0, \qquad \lambda_6^f \leq \frac1{2\pi\lambda_F}.
}
The free energy is
\bel{
  F_{F\pm\pm}\^{crit} = -\frac{N_F V_2 T^2}{\lambda_F} \F \left(\mu_{F\pm\pm},\, \trm{sgn}(\lambda_F)\trm{sgn}\left[\lambda_F\pm 1 \pm \hat\lambda_F\right] e^{i \pi \lambda_F}\right),
}
with
\bel{\label{gap Fcrit}
  \mu_{F\pm\pm}\left(\lambda_F \pm 1 \pm\hat\lambda_F \right) = \frac2\pi \trm{Im\,Li}_2 \left(\trm{sgn}(\lambda_F)\trm{sgn}\left[\lambda_F\pm 1 \pm \hat\lambda_F\right] e^{-\mu_{F\pm\pm} + i \pi \lambda_F}\right).
}
There are now four saddle points, indexed by two independent signs that can enter the free energy (the second sign is the sign of $\sigma$ at the saddle point). Take $\lambda_F > 0$ to simplify the analysis; analogous results hold for $\lambda_F < 0$. Numerical evaluation shows that $F_{F++}\^{crit} < 0$ is the dominant saddle with $\avg a_{F++}\^{crit} = 0$ at all couplings, and conversely $F_{F--}\^{crit} > 0$ is the least dominant saddle with $\avg a_{F--}\^{crit} = \pi$. The other two saddle points display the same type of phase transition seen in the regular bosonic theory; at $\hat \lambda_F = 1 - \lambda_F$ the free energy $F_{-+}\^{crit}$ changes sign from positive to negative, and at $\hat \lambda_F = 1 + \lambda_F$ the free energy $F_{+-}\^{crit}$ changes sign from negative to positive. The critical lines of the two saddles that display the phase transition are thus found to be at
\bel{
  \lambda_6^f(\lambda_F) = \frac1{2\pi\lambda_F} \left(1 - \frac1{(1 \pm \lambda_F)^2} \right).
}
As before, the gap diverges on both of these lines of ``topological points.'' These results can all be seen on Fig.~\ref{fig FFCrit}.

What happens to the bosonization between regular bosons and critical fermions?  Consider the gap equations \eqref{gap Breg} and \eqref{gap Fcrit}. Demanding that the thermal mass and the clumping point are preserved by bosonization yields $\trm{sgn}(\lambda_F) = -\trm{sgn}(\lambda_B)$ and $\lambda_B(1 \pm 2/\hat\lambda_B) = \lambda_F \pm 1 \pm \hat\lambda_F$. The first condition is already in the bosonization rules \eqref{bosonization}, and the second one, at $\lambda_F \geq 0$ and $\lambda_B = -|\lambda_B| \leq 0$, can be rewritten as
\bel{
  \mp 2 \frac{|\lambda_B|}{\hat \lambda_B} = (1 \pm 1) \pm \hat \lambda_F.
}
This equation requires care. Each sign is chosen independently.  The l.h.s.~corresponds to the ``$\pm$'' saddle of the regular boson, and the r.h.s.~corresponds to the ``$\pm\pm$'' saddle of the critical fermion. Studying the choices of signs that can solve this equation, we find that only two fermionic saddles can be bosonized.  The upshot is that the ``$\pm$'' regular boson is dual to the ``$-\mp$'' saddle of the critical fermion. Both of these fermionic saddles are subdominant. Note that the ``$+$'' boson saddle has no phase transitions, just like the ``$--$'' fermion saddle. On the other hand, the ``$-$'' boson saddle and the ``$-+$'' fermion saddle both display a phase transition in the clumping position; for the boson it happens at $\hat\lambda_B = 2$, and for the fermion it happens at $\hat\lambda_F = 1 - \lambda_F$. The topological points are, of course, related by the above duality between marginal deformations, which can be recorded in the simple form
\bel{
  2\frac{|\lambda_B|}{\hat \lambda_B} = \hat\lambda_F.
}
The same map appears in \cite{Aharony:2012ns}; we conclude that all bosonization rules are the same as in the ordinary matter theories, with the all-important proviso that dominant saddles in one theory are mapped to sub-dominant saddles in the other theory, and that there are saddles which are not dualized to anything.

\section{Discussion}

Let us summarize our findings. We have studied CS-ghost theories with $N$ matter fields and found that, on $\R^3$ and to all orders in $N$ and the couplings, they are related by $N \rightarrow -N$ to appropriate CS-matter theories with ordinary matter. This implies that bosonization on $\R^3$ holds for these non-unitary models. These results are encapsulated in Figure \ref{fig arrowsdiag}. At very high temperatures, we have shown that $N \rightarrow -N$ and bosonization fail in the large $N$ limit. For both ordinary and ghost matter with six-point couplings, we have found that the line of spontaneous conformal symmetry breaking points in coupling space, previously seen at zero temperature, at high temperature becomes a line of points at which the theory is topological and across which the Polyakov loop eigenvalue clumping position jumps between $0$ and $\pi$.

There are several directions for further research, as has been indicated throughout the text:
\begin{enumerate}
  \item It would be of interest to examine the singlet content of all the models presented here and to study their mutual mappings. In particular, it would be interesting to understand whether baryon operators exist in these theories and if so, how to interpret them from the point of view of the bulk de Sitter dual; perhaps they are related to the failure of bosonization at high temperatures.
  \item The full steepest descent analysis of the high-$T$ phase of these theories has not yet been performed. As we briefly discussed, at present this task presents technical difficulties, but completing it is necessary in order to be completely confident in our large $N$ results.
  \item Intermediate temperatures likely contain very rich phase structures. The analysis initiated in \cite{Jain:2013py} should largely carry over to the CS-ghost case. It would be fascinating to understand the phase structure of both kinds of models at all temperatures. Doing so would also afford us a better understanding of the fate of the conformal symmetry breaking transition at high temperatures.
  \item It would be very interesting to study the partition functions as functions of $\lambda$. According to the dS/CFT dictionary, these are computing pieces of the late time Hartle-Hawking wavefunction. We hope to understand how the issues on the normalizeability of the wavefunction (on an $S^3$) raised in \cite{Anninos:2012ft} for $\lambda=0$ are affected by the inclusion of the Chern-Simons sector.
\end{enumerate}

\section*{Acknowledgements}

It is a great pleasure to thank Shamik Banerjee, Frederik Denef, Guy Gur-Ari, Sean Hartnoll, Chao-Ming Jian, George Konstantinidis, Ran Yacoby, Steve Shenker and Andy Strominger. The work of \DJ.R. is supported by an NSF Graduate Fellowship. R.M. is supported by a Gerhard Casper Stanford Graduate Fellowship. E.S. is supported in part by the Oskar Heil Stanford Graduate Fellowship and NSF Grant PHY-0756174. This work has been partially funded by DOE grant DE-FG02-91ER40654.

\appendix

\section{Symplectic groups and vector models with anticommuting bosons} \label{Appendix SpN}

There are multiple ``symplectic groups.'' Two non-compact ones are of interest to us, $Sp(2N,\R)$ and $Sp(2N, \C)$. These are groups of $2N\times 2N$ (respectively) real- and complex-valued matrices $G$ that preserve the symplectic structure $\Omega = \left[^{\phantom{-\1_N} \1_N}_{-\1_N}\right]$, i.e.~that satisfy $G^T \Omega G = \Omega$.

The real symplectic group $Sp(2N, \R)$ has $2N^2 + N$ generators. In the fundamental representation, these are $2N \times 2N$ matrices
\bel{
  T = \bmat{A}{B}{C}{-A^T},
}
where $A$ is an arbitrary $N\times N$ matrix while $B$ and $C$ are symmetric $N \times N$ matrices. The group element generated by these $T$'s is
\bel{
  G = e^{\theta^a T^a}.
}
The basis can be chosen such that there are $N^2 + N$ purely real, symmetric generators  $T^a\_{S}$, and $N^2$ purely real, antisymmetric generators $T^a\_{AS}$,\footnote{In this notation, the index $a$ runs over all $2N^2 + N$ generators, generically denoted $T^a$. It is understood that $T^a\_{S} = 0$ for those $a$ for which $T^a$ is not symmetric; an analogous rule holds for $T^a\_{AS}$.} with the Cartan-Killing metric in a Minkowski form,
\bel{\label{def eta}
  \Tr\left(T^a T^b\right) = \eta^{ab},
}
with $N^2 + N$ positive entries corresponding to the symmetric generators and $N^2$ negative entries corresponding to the antisymmetric generators.

The complex symplectic group $Sp(2N, \C)$ is generated by $4N^2 + 2N$ matrices. Its Lie manifold is a complexification of the $Sp(2N, \R)$ one. With the same generators as before, a generic group element $G$ can be written as
\bel{
  G = e^{\theta^a T^a + \~\theta^a \~T^a},
}
with $\~T^a = iT^a$. It is of note that the choice of generators $T^a$ above does not lead to a diagonal Cartan-Killing metric $\Tr\left( T^aT^b\right)$. This metric can be diagonalized in a ``light cone'' basis of linear combinations of the real and imaginary versions of $T^a\_S$ and $T^a\_{AS}$, but this is not necessary for our purposes.

One often needs to use the largest compact subgroups of the two groups introduced above. For $Sp(2N, \R)$, this is the group of symplectic orthogonal matrices, generated by the $N^2$ antisymmetric generators $T^a\_{AS}$. This group is isomorphic to $U(N)$. For $Sp(2N, \C)$, the largest compact subgroup is the group of symplectic unitary matrices, generated by the $2N^2 + N$ anti-Hermitian generators (the $T\_{AS}$'s and the $\~T\_S$'s). This group also can be realized as an analytic continuation $T^a\_S \rightarrow iT^a\_S$ of $Sp(2N, \R)$, \emph{or} as a group $U(N, \mathbb H)$ of unitary matrices over the quaternions, and it is commonly denoted $USp(N)$, $Sp(N)$, or $Sp(2N)$. We will use $USp(N)$.

Which Lagrangians with anticommuting scalar matter in the fundamental representation possess these symplectic symmetries? The answer must be determined by studying the bilinear structures that are invariant under each group action.\footnote{This discussion is similar in spirit to one found in \cite{Giombi:2013fka}, but there the focus was on the ``ordinary statistics'' theories.} The $Sp(2N, \R)$ bosonic vector model has a single invariant, $\Omega_{ij} \chi_i \chi_j$, and the Lagrangian must take the form
\bel{\label{Sp2NRmodel}
  \L_{Sp(2N, \R)} = \frac12 \Omega_{ij} \del_\mu \chi_i \del^\mu \chi_j + V\left(\frac12\Omega_{ij} \chi_i \chi_j\right).
}
The invariants of the largest compact subgroup of $Sp(2N, \R)$ are all the invariants of $Sp(2N, \R)$ and all the invariants of $O(2N)$; thus we may also use both $\Omega_{ij} \chi_i \chi_j$ and $\chi_i \chi_i$ to construct invariant operators. However, due to the anticommutativity of the matter, the latter term is always identically zero, and the ``compactification'' of the $Sp(2N, \R)$ model must be a theory with the same Lagrangian as before, namely \eqref{Sp2NRmodel}. By setting $\chi_{i + N} \equiv \bar \chi_i$ for $i \leq N$, this Lagrangian takes on a form with manifest $U(N)$ symmetry,
\bel{\label{UNmodel}
  \L_{U(N)} = \del_\mu \bar \chi_i \del^\mu \chi_i + V\left(\bar\chi_i \chi_i\right),
  \quad 1\leq i \leq N.
}
Thus, on the surface, the $Sp(2N, \R)$ and $U(N)$ ghost boson models have the same Lagrangian, rewritten in \eqref{Sp2NRmodel} and \eqref{UNmodel} to make the relevant symmetries manifest. These models, however, do \emph{not} have the same singlet operator content. The $Sp(2N, \R)$ singlet model has only even-spin conserved currents, while the $U(N)$ singlet model has currents of all spins. For instance, the spin-one current $\bar\chi_i \del_\mu \chi_i - (\del_\mu\bar\chi_i) \chi_i$ is a singlet under $U(N)$ but not under $Sp(2N, \R)$, while spin-zero and spin-two currents, $\bar\chi_i \chi_i$ and $\del_{(\mu}\bar\chi_i \del_{\nu)} \chi_i - \delta_{\mu\nu} \del_\lambda \bar\chi_i \del^\lambda \chi_i$, are singlets of both groups.\footnote{The group action of $Sp(2N, \R)$ becomes natural once the model is de-complexified via $\chi_{i + N} \equiv \bar \chi_i$. De-complexifying the spin-one current, we see that it can be written as $\chi_{i + N}\del_\mu \chi_i - (\del_\mu \chi_{i + N}) \chi_i = \chi^T \left[^{\phantom{\1_N} \1_N}_{\1_N}\right] \del_\mu\chi$. This is not invariant under arbitrary $Sp(2N, \R)$ transformations.}

The $Sp(2N, \C)$ and $USp(N)$ models are richer. The non-compact $Sp(2N, \C)$ model has a single \emph{complex} invariant, $\Omega_{ij} \chi_i \chi_j$. Here we define $\chi_i\+ \equiv \bar\chi_i$, and refer to quantities obeying $A\+ = A$ as real or Hermitian. The most general Hermitian Lagrangian is
\bel{\label{Sp2NCmodel}
  \L_{Sp(2N,\C)} = \frac12 \Omega_{ij} \left( \del_\mu\chi_i \del^\mu \chi_j + \del_\mu\bar\chi_i \del^\mu \bar\chi_j \right) + V\left( \frac12\Omega_{ij}\chi_i\chi_j,\, \frac12 \Omega_{ij} \bar\chi_i \bar\chi_j \right),
}
where $(\chi_i, \bar\chi_i)$ are $2N$ pairs of conjugate complex numbers that are treated as independent, and $V(x,y)$ is a function obeying appropriate reality conditions. This model has no odd-spin conserved current singlets, but it has two copies of each $Sp(2N, \R)$ conserved current singlet (or, in other words, each current singlet has two components). The $USp(N)$ model has an additional non-trivial invariant, $\bar \chi_i \chi_i$, and hence the general $USp(N)$ anticommuting Lagrangian is
\bel{\label{USpNmodel}
  \L_{USp(N)} =  Z \del_\mu \bar\chi_i \del^\mu \chi_i + \frac12 \Omega_{ij} \left( \del_\mu\chi_i \del^\mu \chi_j + \del_\mu\bar\chi_i \del^\mu \bar\chi_j \right) + V\left(\bar \chi_i \chi_i,\, \frac12\Omega_{ij}\chi_i\chi_j,\, \frac12 \Omega_{ij} \bar\chi_i \bar\chi_j \right).
}
The Lagrangian for the compact subgroup, \eqref{USpNmodel}, is not the same as that of the original non-compact group, \eqref{Sp2NCmodel}. There are additional terms that can be written in a general $USp(N)$ model, such as the marginal term $(\bar\chi_i\chi_i)^2 \Omega_{jk} (\chi_j\chi_k + \bar\chi_j\bar\chi_k)$. However, like in the previous case, moving to the compact subgroup of $Sp(2N, \C)$ increases the number of conserved currents. Odd-spin currents are also present, such as the spin-one, $Z \left(\bar\chi_i \del_\mu \chi_i - (\del_\mu \bar \chi_i)\chi_i \right)  + \Omega_{ij}(\chi_i \del_\mu \chi_j - \bar\chi_i \del_\mu \bar \chi_j)$. Studying all these operators in a unified fashion will be the subject of a forthcoming publication.

The complex matter models can be projected onto either the $\chi$ or the $\bar\chi$ sector by simply setting $\bar \chi = 0$ or $\chi = 0$, respectively. Such a Lagrangian takes the same form as \eqref{Sp2NRmodel}, but with non-Hermitian fields:
\bel{\label{USpNProjmodel}
  \L_{USp(N)}\^{proj} = \frac12 \Omega_{ij} \del_\mu \chi_i \del^\mu \chi_j + V\left(\frac12\Omega_{ij} \chi_i \chi_j\right).
}
This ruins the hermiticity of the Lagrangian and removes some conserved current singlets from the spectrum. In particular, the $USp(N)$ model restricted to the $\chi$ sector contains only a single operator of each even spin.

We close this overview with a brief remark on the $U(N, \mathbb H)$ group mentioned above. In the case of real symplectic groups, we have shown that pairs of real Grassmann fields can be packaged into a single complex field so as to make the $U(N)$ symmetry apparent. The same trick can be applied to the complex case, but then we need to package two complex numbers into a single quaternion. There are multiple ways to represent this quaternion; a standard way is to take two complex numbers $z$ and $w$ and define $h =  \left[^{\ z\ \ w}_{-\bar w\ \bar z}\right]$. Each such quaternion can be decomposed as $h = h_\mu \sigma^\mu$, where $\mu = 0, \ldots, 3$, $h_\mu$ are real Grassmann numbers, and $\sigma^\mu = (\1_2, i \boldsymbol\sigma)$ is, up to a factor of $i$, the standard four-vector built of Pauli matrices. This way, the unitary $Sp(2N, \C)$ actions can be represented as unitary quaternion-valued $N \times N$ matrices acting on $h$. The single invariant of this group is $h_i\+ h_i$ with conjugation acting as usual Hermitian conjugation on the Pauli matrices. Decomposing this product into Pauli matrices shows that it has three independent components, one for each Pauli matrix, and these precisely match with the three $USp(N)$ invariants identified in the previous passage.

\section{Feynman diagrams in Landau gauge on $\mathbb{R}^3$}
\label{app:diagrams}

In this appendix, we explicitly compute all the relevant Feynman diagrams
in four models, all with bosonic matter: $O(N)$, $U(N)$, $Sp(N,\mathbb{R})$ with ghost matter, and $U(N)$ with ghost matter. The $O(N)$ case was done earlier in \cite{Aharony:2011jz}, whose conventions we follow.

We work in three Euclidean dimensions, and weight in the path integral is always $e^{-S}$. The action is, as usual, $S = \int \d^3x\, \mathcal{L}$. We will write some of the terms in the Lagrangian in momentum space since it makes sign conventions clear. For momenta, arrows going \emph{out} of a vertex are treated as positive.

The cubic coupling in the Chern-Simons action has the same for all the groups and we write
\begin{align}
\mathcal{L} &\supset - \frac{ig}{6} \epsilon_{\mu\nu\lambda} f^{abc} A^a_{\mu}A^b_{\nu}A^c_{\lambda}.
\end{align}
We work in the Landau gauge $\partial_\mu A_\mu = 0$ always. The gauge field propagator is
\begin{align}
\langle A_\mu (q) A_\nu (-q) \rangle = -\epsilon_{\mu\nu a} \frac{q^a}{q^2}.
\end{align}

\begin{center}
\vspace{-0.2in}
\includegraphics[width=1.5in]{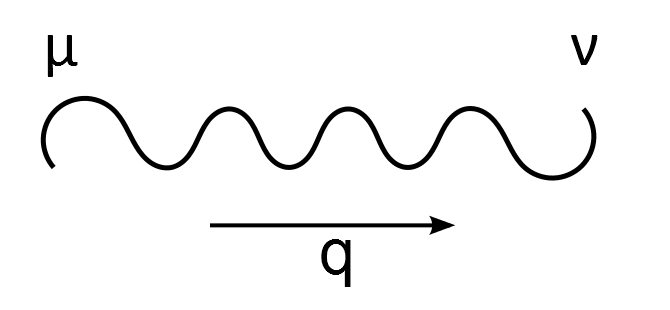}
\end{center}
\vspace{-0.5in}
\subsection{$O(N)$}
The matter part of the Lagrangian is given by
\begin{align}
\mathcal{L} &\supset \frac{1}{2} (D \phi_i)(D \phi_i)
+ \frac{g_6}{48} (\phi_i\phi_i)^3, \quad
D\phi_i = \partial \phi_i + g\, T^a_{ij}A^a \phi_j.
\end{align}
The matrices $T^a$, the generators of $O(N)$, are antisymmetric.
The kinetic term contains the following cubic and quartic couplings:
\begin{align}
\mathcal{L} &\supset -\frac{ig}{2}(q+r)^\mu T^a_{ij} A^a_\mu(p) \phi_i(-q)\phi_j(r) (2\pi)^3\delta^3(p-q+r), \\
\mathcal{L} &\supset -\frac{g^2}{4} \{T^a,T^b\}_{ij} A^a_\mu A^b_\mu \phi_i \phi_j.
\end{align}
In computing the momentum integrals, we take only the $(q+r)^\mu$ factor from the three point vertex, the rest of the factors are adjusted in the combinatorics.
The six point coupling is normalized such that the tree-level six point function is
\begin{align}
\langle \phi_{i_1}\ldots \phi_{i_6} \rangle_{\text{tree}}
= -g_6 \left( \delta_{i_1i_2}\delta_{i_3i_4}\delta_{i_5i_6} + 14
\text{ permutations} \right).
\end{align}

\subsection{$Sp(N,\mathbb{R})$}
We assume that $N$ is an even integer.
We denote by $\Omega$ the canonical antisymmetric matrix. By the definition of the $Sp(N,\mathbb{R})$ group, the generators $T^a$ all satisfy $(T^a)^T = \Omega T^a \Omega$.
For convenience, we define the matrices
$S^a = \Omega T^a$,
which are symmetric.
The relevant Fierz identity needed to compute Feynman diagrams is
\begin{align}
  \eta_{ab} T^a_{ij} T^b_{kl} = \frac12 \left( \delta_{il} \delta_{jk} - \Omega_{ik} \Omega_{jl} \right).
\end{align}

The matter part of the Lagrangian is
\begin{align}
\mathcal{L} &\supset \frac{1}{2} \Omega_{ij} (D \chi_i)(D \chi_j)
+\frac{g_6}{48} \left(\Omega_{ij}\chi_i\chi_j\right)^3, \quad
D\chi_i = \partial \chi_i + g\, T^a_{ij}A^a \chi_j.
\end{align}
The cubic and quartic couplings are
\begin{align}
\mathcal{L} &\supset -\frac{i g}{2} (q+r)^\mu S^a_{ij} A^a_\mu(p)
\chi_i(-q) \chi_j(r)
(2\pi)^3 \delta^3(p-q+r),\\
\mathcal{L} &\supset -\frac{g^2}{4} A^a_\mu A^b_\mu [\Omega \{T^a,T^b\}]_{ij} \chi_i\chi_j.
\end{align}
The propagator and the tree-level six-point vertex are as follows:
\begin{align}
\langle\chi_i(p) \chi_j(-p) \rangle &= - \frac{\Omega_{ij}}{p^2},\\
\langle\chi_{i_1}\ldots \chi_{i_6} \rangle_{\text{tree}} &= g_6
\left( \Omega_{i_1i_2}\Omega_{i_3i_4}\Omega_{i_5i_6}
\pm 14 \text{ permutations} \right).
\end{align}

\subsection{$U(N)$}
The group $U(N)$ contains  $SU(N)$ and $U(1)$ subgroups. In general, the coupling constants for these subgroups could be different. We work with the case when the two couplings are equal. The other cases are qualitatively the same.
The required interaction terms in the Lagrangian are given below:
\begin{align}
\mathcal{L} &\supset -ig(q+r)^\mu T^{a}_{ij}
A_\mu(p) (\phi_i(q))^\dagger \phi_j(r) (2\pi)^3\delta^3(p-q+r),\\
\mathcal{L} &\supset -\frac{g^2}{2} \{T^a,T^b\}_{ij} A^a_\mu A^b_\mu\phi_i^\dagger \phi_j, \\
\mathcal{L} &\supset \frac{g_6}{6} (\phi^\dagger \phi)^3.
\end{align}
To contract gluons we use the Fierz identity
\bel{\label{UNFierz}
  T^a_{ij} T^a_{kl} = \delta_{il} \delta_{jk},
}
and the propagator and the tree-level six-point amplitude are
\begin{align}
\langle \phi_{i_1} \phi^\dagger_{i_2} \rangle &= \frac{1}{p^2} \delta_{i_1i_2},\\
\langle
\phi^\dagger_{i_1} \phi_{i_2}\phi^\dagger_{i_3} \phi_{i_4}\phi^\dagger_{i_5} \phi_{i_6}
\rangle
&= -g (\delta_{i_1i_2}\delta_{i_3i_4}\delta_{i_5i_6} + 5 \text{ permutations}).
\end{align}
The only thing that changes for the $U(N)$ theory with ghost matter is that we have to be careful about the ordering of $\phi$ and $\phi^\dagger$ in the propagator, and that the
tree level six point function is modified to incorporate the correct signs in the tensor factor:
\begin{align}
\langle
\phi^\dagger_{i_1} \phi_{i_2}
\phi^\dagger_{i_3} \phi_{i_4}
\phi^\dagger_{i_5} \phi_{i_6}
\rangle
= -g (\delta_{i_1i_2}\delta_{i_3i_4}\delta_{i_5i_6} \pm 5 \text{ permutations})
\end{align}

\subsection{Evaluation of the Diagrams}

The momentum integrals for these diagrams are found to be
\begin{align}
(A1) = \frac{1}{32 \pi^2 \epsilon}, \,
(A2) = \frac{1}{32\pi^2 \epsilon}, \,
(A3) = \frac{-1}{32 \pi^2 \epsilon}, \,
(A4) = \frac{-1}{16 \pi^2 \epsilon}, \nonumber \\
(A5) = \frac{-1}{32 \pi^2 \epsilon}, \,
(A6) = \frac{-3}{64 \pi^2 \epsilon}, \,
(A7) = \frac{3}{64 \pi^2 \epsilon}, \,
(A8) = \frac{1}{32 \pi^2 \epsilon}, \\
(B1) = \frac{1}{24 \pi^2 \epsilon}, \,
(B2) = \frac{1}{96 \pi^2 \epsilon}, \,
(B3) = \frac{1}{12 \pi^2 \epsilon}, \,
(B4) = \frac{1}{24 \pi^2 \epsilon} . \nonumber
\end{align}

\begin{table}
\begin{center}
\begin{tabular}{|c|c|c|c|}
\hline
Diagram &  & $O(N)$ & $Sp(N, \mathbb{R})$ ghost-like  \\
\hline
A1& $g^4g_6$  & $(-N^2 - 7N + 8)3/128$   & $(-N^2 + 7N + 8)3/128$\\
A2& $g^8$     & $(N^2 + N - 2)3/64$        & $(N^2 - N - 2)3/64$\\
A3& $g^8$     & $(-N^2 + 3N - 2)3/64$      &  $-(N^2 + 3N + 2)3/64$\\
A4& $g^4g_6$  & $(- N + 1)9/32$            & $(N + 1)9/32$\\
A5& $g^8$     & $(N - 1)3/64$              & $-(N + 1)3/64$\\
A6& $g^8$     & $0$                        & $0$ \\
A7& $g^8$     & $(N-1)9/64$                & $-(N+1)9/64$\\
A8& $g_6^2$   & $(3N+22)/32$               & $(-3N+22)/32$\\
\hline
B1& $g^4$     & $(-N^2+3N-2)/96$           & $-(N^2+3N+2)/96$\\
B2& $g^4$     & $(N^2-N)/384$              & $(N^2+N)/384$\\
B3& $g^4$     & $(N-1)/48$                 & $-(N+1)/48$\\
B4& $g^4$     & $(N-1)/96$                 & $-(N+1)/96$\\
\hline
\end{tabular}
\end{center}
\caption{\small Diagrams for $O(N)$ and $Sp(N, \mathbb{R})$. The value of any diagram is equal to $1/(\pi^2\epsilon)$ times the product of the coupling constants in the second column and the numerical factor in the third or the fourth column, depending on the theory.}
\label{tab:onspn}
\end{table}
\noindent For the $O(N)$ and the $Sp(N,\mathbb{R})$ theories,
the diagrams evaluate to the results given in Table \ref{tab:onspn} above.
For the $U(N)$ theories,
the diagrams evaluate to the results given in Table \ref{tab:unuminusn} below.
The entries in the table are to be understood as follows, with all the external
legs truncated:
\begin{align}
\langle \phi_i \phi_j \rangle &=
\delta_{ij} \frac{p^2}{\pi^2 \epsilon} \times (\text{entry in table}), \\
\langle \chi_i \chi_j \rangle &=
\Omega_{ij} \frac{p^2}{\pi^2 \epsilon} \times (\text{entry in table}),\\
\langle \phi_i \phi_j^\dagger \rangle &=
\delta_{ij} \frac{p^2}{\pi^2 \epsilon} \times (\text{entry in table}),\\
\langle \phi_{i_1}\ldots \phi_{i_6} \rangle
&= \left( \delta_{i_1i_2}\delta_{i_3i_4}\delta_{i_5i_6} + 14
\text{ permutations} \right) \times (\text{entry in table}) \times \frac{1}{\pi^2\epsilon}, \\
\langle\chi_{i_1}\ldots \chi_{i_6} \rangle &=
-\left( \Omega_{i_1i_2}\Omega_{i_3i_4}\Omega_{i_5i_6}
\pm 14 \text{ permutations} \right) \times (\text{entry in table}) \times \frac{1}{\pi^2\epsilon},  \\
\langle \phi^\dagger_{i_1} \phi_{i_2}\phi^\dagger_{i_3} \phi_{i_4}\phi^\dagger_{i_5} \phi_{i_6} \rangle &= (\delta_{i_1i_2}\delta_{i_3i_4}\delta_{i_5i_6} + 5 \text{ permutations}) \times (\text{entry in table}) \times \frac{1}{\pi^2\epsilon}.
\end{align}
For the $U(N)$ case with ghost matter,
the last equation is modified only by the replacement of the
tensor structure with appropriate minus signs.

\begin{table}
\begin{center}
\begin{tabular}{|c|c|c|c|}
\hline
Diagram &  & $U(N)$ & $U(N)$ ghost-like  \\
\hline
A1& $g^4g_6$  & $-(N^2+4N+5)3/32$ & $-(N^2-4N+5)3/32$\\
A2& $g^8$     & $(N^2 + 5N + 10)3/16$ &  $(N^2 - 5N + 10)3/16$\\
A3& $g^8$     & $-(N^2 + 5N - 6)3/16$ & $-(N^2 - 5N - 6)3/16$\\
A4& $g^4g_6$  & $-(2N+1)3/8$ & $(2N-1)3/8$\\
A5& $g^8$     & $(N+3)3/16$ & $(-N+3)3/16$\\
A6& $g^8$     & $(-N+1)9/32$ & $(N+1)9/32$\\
A7& $g^8$     & $(3N+5)9/32$ & $(-3N+5)9/32$ \\
A8& $g_6^2$   & $(3N+11)/16$ & $(-3N+11)/16$ \\
\hline
B1& $g^4$     & $(-N^2+1)/24$ & $(-N^2+1)/24$\\
B2& $g^4$     & $(N^2+1)/96$ &   $(N^2+1)/96$\\
B3& $g^4$     & $1/12$ & $1/12$  \\
B4& $g^4$     & $N/24$ & $-N/24$  \\
\hline
\end{tabular}
\end{center}
\caption{\small Diagrams for the bosonic $U(N)$ theories. The value of any diagram is equal to $1/(\pi^2\epsilon)$ times the product of the coupling constants in the second column and the numerical factor in the third or the fourth column, depending on the theory.}
\label{tab:unuminusn}
\end{table}


Define $Z = 1 + B$ and
$g_6Z_6 = g_6+ A$. The quantities $A$ and $B$ are simply the sum of all the
$A$ diagrams and $B$ diagrams given in the table, respectively, with an
additional factor of $1/\pi^2$.
The $\beta$-function is given by
\begin{align}
\mu\frac{\d}{\d\mu}g_6 = 2A - 6B g_6.
\end{align}
The appropriate couplings that are kept fixed in the large $N$ limit are:
$\lambda = g^2 N$,  and $\lambda_6 = g_6 N^2$.
Adding up all the diagrams to calculate the quantities $A$ and $B$ gives us the appropriate beta functions.

For $O(N)$ and $Sp(N, \R)$, we find the $\beta$-function
\begin{align}
\beta_{\lambda_6} = \frac{1}{16\pi^2 N^2} \left(
12 \lambda^4 (\pm N-1) - 20 \lambda^2 \lambda_6 (\pm N-1) + \lambda_6^2(\pm3N+22)
\right),
\end{align}
where the upper sign pertains to $O(N)$ with commuting bosons and the lower sign to $Sp(N,\mathbb{R})$ with anticommuting bosons.
For the two $U(N)$ theories, we find
\begin{align}
\beta_{\lambda_6} = \frac{1}{16\pi^2N^2}\left(
3\lambda^4 (\pm 11N+53) -40\lambda^2 \lambda_6(\pm N+1) + 2\lambda_6^2(\pm 3N+11)
\right),
\end{align}
where the upper sign pertains to $U(N)$ with
commuting bosons and the lower sign to $U(N)$ with anticommuting bosons. These results are discussed in Section \ref{section: landau}.

\section{Integrating out the CS sector in light-cone gauge}\label{Appendix LCgauge}

In this Appendix, we implement the light-cone gauge for $U(N)$ bosonic theories in detail and show how the reduce the path integral \eqref{def ZB} to a vector model with non-local couplings.  Here and in the next Appendix, we follow the conventions of \cite{Aharony:2012ns}, which at some places are different than those of \cite{Aharony:2011jz}. The gauge-fixed action is, in gory detail,
\algnl{\notag
  S
  &= S\_{CS} + \int\! \d^3x\! \left[\del_\mu \bar \chi_i + (A_\mu)^*_{ij} \bar \chi_j \right]\!\!\left[\del_\mu \chi_i + (A_\mu)_{ij} \chi_j \right] + \!\int\! \d^3x\! \left( \sigma \bar \chi_i \chi_i + \frac{\lambda_6^b(\bar\chi_i \chi_i)^3}{3! N^2} \right)\\ \notag
  &= \int \pdnvol[3] \left[p_s^2 + p_3^2 + \sigma\right] \bar \chi_i(-p) \chi_i(p) - \frac{ik}{4\pi} \int \pdnvol[3] p_- A_+^a(p) A_3^a(-p)\\ \notag
  &\quad - i\int_{p,\,q,\,s} (2\pi)^3 \delta^3(-p + q + s) \bar\chi_i(-p) T_{ij}^a \chi_j(s) \left[(p + s)_- A_+^a(q) + (p + s)_3 A_3^a(q)\right]  \\ \notag
  &\quad - \int_{p,\,q,\,s,\,t} (2\pi)^3 \delta^3(-p + q + s + t) \bar \chi_i(-p)T^a_{ij}T^b_{jk}\chi_k(t)  A_3^a(q) A_3^b(s)  \\ \label{S full R3}
  &\quad + \frac{\lambda_6^b}{3! N^2}\int_{p_1,\, \ldots,\, p_6} (2\pi)^3 \delta\left(\sum p_i\right) \bar\chi_i(p_1) \chi_i(p_2)\bar\chi_j(p_3) \chi_j(p_4)\bar\chi_k(p_5) \chi_k(p_6).
}

Integrating out $A_+$ (now a mere Lagrange multiplier) from the above action enforces the equality
\bel{
  A_3^a(q) = \frac{4\pi}{kq_-} \int_p \bar\chi_i(-p) T_{ij}^a \chi_j(p + q) (2p + q)_-,
}
and plugging this back into the action and using Fierz identities $T^a_{ij} T^a_{kl} = -\frac12 \delta_{il}\delta_{jk}$ gives a vector model in a background field,
\algnl{\notag
  Z_B
  &= \int[\d\chi] \exp\Biggr\{-\int_p \bar\chi_i(-p) \chi_i(p) \left[p_s^2 + p_3^2 + \sigma\right] - N \int_{P,\,p\,q} \chi^2(P, p) \chi^2(-P, q) \frac{2\pi i N}{k} C_1(P, p, q)\\ \label{ZB}
  &\hspace{-2em}- N \int_{P,\,Q,\,R,\,p,\,q,\,r} \hspace{-3em}(2\pi)^3\delta^3(P + Q + R)\, \chi^2(P, p)\, \chi^2(Q, q)\, \chi^2(R, r) \left(\frac{4\pi^2 N^2}{k^2} C_2(P, Q, R, p, q, r) + \frac{\lambda^b_6}{3!}\right)\Biggr\},
}
We use notation similar to the one found in \cite{Jain:2012qi}, and in the above equation we let
\algnl{
  \chi^2(P, p) &\equiv \frac1N \bar\chi_i \left(\frac P2 - p\right) \chi_i \left(\frac P2 + p\right),\\
  C_1(P, p, q) &\equiv \frac{(P + p + q)_- (-P + p + q)_3}{(p - q)_-},\\
  C_2(P, Q, R, p, q, r) &\equiv \frac{(P - R + 2p + 2r)_- (R - Q + 2q + 2r)_-}{(P + R + 2p - 2r)_- (R + Q + 2r - 2q)_-}.
}
The $U(N)$ regular boson partition function differs from this one only by the sign of the $C_1$ coefficient.

The restricted $USp(N)$ ghost boson and $O(2N)$ regular boson partition function can be written in essentially the same form as \eqref{ZB} above. The only change is that now we must define $\chi^2(P, p) \equiv \frac1N \Omega_{ij} \chi_i\left(\frac P2 - p\right) \chi_j \left(\frac P2 + p\right)$ for ghosts and $\phi^2\equiv \frac1N \phi_i\left(\frac P2 - p\right) \phi_i \left(\frac P2 + p\right)$ for the regular bosons, and the coefficients $C_{1/2}$ differ by various factors of two from the $U(N)$ ones. These two partition functions are related by $\chi^2 \leftrightarrow \phi^2$ and a change of sign of $C_1$, just as above.

\section{Free energy at high temperatures}\label{Appendix HighT}

\subsection{Bosonic models}
The approach of Appendix \ref{Appendix LCgauge} can be used to study large-$N$ $U(N)$ CS-ghost theories on $S^1_\beta \times S^2$. Other than changing the background manifold to contain a thermal circle, the only other new element is the inclusion of a background field $\A_\mu$ that should be added to each $A_\mu$ in \eqref{S full R3} \cite{Aharony:2012ns}. The holonomy $(\A_\mu)_{ij} = \delta^3_\mu \A_{ij} \equiv -i \delta_\mu^3 \delta_{ij} a/\beta$ is not integrated out; it is gauge-fixed to a diagonal form and is taken not to have quantum fluctuations in our high-temperature regime. Instead of arranging the holonomy eigenvalues to lie equally spaced in an interval of width $2\pi|\lambda|$ and centered at $\avg a$, we treat $a$ as a random variable uniformly distributed in this interval, and any momentum integral is understood to also average over this random variable, as long as it is present in the integrand. The resulting vector model is
\algnl{\notag
  Z_B
  &= \int[\d\chi] \exp\Biggr\{-\int_p \bar\chi_i(-p) \chi_i(p) \left[p_s^2 + \left(p_3 - \frac{a}\beta\right)^2 + \sigma\right]  \\
  &\hspace{-2em}- N \int_{P,\,Q,\,R,\,p,\,q,\,r} \hspace{-3em}(2\pi)^3\delta^3(P + Q + R)\, \chi^2(P, p)\, \chi^2(Q, q)\, \chi^2(R, r) \left(\frac{4\pi^2 N^2}{k^2} C_2(P, Q, R, p, q, r) + \frac{\lambda^b_6}{3!}\right)\Biggr\},
}
with the same notation as before. In each integral it is understood that
\bel{
  \int \pdnvol[3] \equiv \frac1\beta \sum_{n \in \Z} \int_{-1/2}^{1/2} \d u \int_0^\infty \frac{p_s \d p_s}{2\pi},
}
with $a \equiv \avg a + 2\pi u |\lambda|$ and $p_3 = \frac{2\pi}\beta\left(n + \frac12\right)$.
The quartic term present in eq.~\eqref{ZB} drops out, as in \cite{Jain:2012qi}.
Next, 
we insert 
\begin{align}
\int [\d\alpha~\d\mu] \exp\left\{-i N \mu(P,p) \left[\alpha(P, p) - \chi^2(P, p)\right]\right\}
\end{align}
into the partition function and obtain the saddle point equations
\algnl{\notag
  i\mu(P, p)
  &\equiv \Sigma(p) (2\pi)^3 \delta^3(P) \\ \label{SD Sigma}
  &= -4\pi^2 \lambda^2 (2\pi)^3 \delta^3(P) \int_{p, q, r}\hspace{-1em} \alpha(q) \alpha(r) \left[C_2(p, q, r) + C_2 (q, p, r) + C_2 (q, r, p) + \frac{\lambda_6^b}{8\pi^2 \lambda^2}\right],
}
and
\bel{\label{SD alpha}
  \alpha(P, p) = (2\pi)^3 \delta^3(P) \alpha(p) = - \frac{(2\pi)^3 \delta^3(P)}{\eps(p) - \Sigma(p)},\quad \eps(p) = p_s^2 + \left(p_3 - \frac{a}\beta\right)^2 + \sigma,
}
where $C_2(p,q,r) \equiv C_2(0,0,0,p,q,r)$. The change of variables $i\mu \equiv \Sigma$ signifies that the integral over $\Sigma$ is done over the imaginary axis.

Now, writing $(2\pi)^3 \delta^3(0) \equiv V_3$, the free energy at the saddle point can be expressed as
\bel{
  F_B \equiv - \log Z_B = - N V_3 \int\pdnvol[3] \left[ \log\big(\eps(p) - \Sigma(p) \big) + \frac23 \frac{\Sigma(p)}{\eps(p) - \Sigma(p)} \right].
}
The shifts in Matsubara frequencies  and the overall minus sign in $F_B$ are the only differences from the commuting case.

Solving the gap equation now amounts to finding saddle points of $F_B$ as a function of $\Sigma(p)$ and $\sigma$ after summing and regulating the above expression. It turns out that at the saddle point $\Sigma(p)$ is independent of momentum and lies on the real axis (recall that the integral over $\Sigma(p)$ runs over the imaginary axis). We will refer to the saddle point value as just $\Sigma$. Setting $\mu_B^2 \equiv \beta^2(\sigma - \Sigma)$ with $\mu_B \geq 0$, 
and doing the Matsubara sum following the renormalization prescription of \cite{Aharony:2012ns}, we get
\algnl{\notag
  \int \pdnvol[3] \log\big(\eps(p) - \Sigma  \big)
  &= \int_{-1/2}^{1/2}\frac{\d u }{2\pi\beta^3} \left\{-\frac{\mu_B^3}3 + \int_{\mu_B}^\infty\!\!\! \d y\, y\,\log\left[1 + e^{-2y} + 2e^{-y} \cos(2\pi |\lambda|u + \avg a) \right]  \right\}\\
  &\hspace{-20ex}= \frac1{2\pi\beta^3} \left\{-\frac{\mu_B^3}3 - \frac1{\pi|\lambda|} \int_{\mu_B}^\infty\!\!\! \d y\, y\left[\trm{Im}\, \trm{Li}_2\left(-e^{-y + i\pi|\lambda| + i \avg a} \right) + \trm{Im}\, \trm{Li}_2\left(-e^{-y + i\pi|\lambda| - i \avg a} \right) \right]\right\}.
}
Here we have used the dilogarithm function $\trm{Li}_2(x) \equiv \sum_{m = 1}^\infty \frac{x^m}{m^2}$. The saddle-point value of the integral of the second term can be found without doing any integrals. The Schwinger-Dyson equation for the thermal mass can be obtained by substituting (\ref{SD alpha}) into (\ref{SD Sigma}):
\bel{
  \Sigma = \sigma - \beta^{-2} \mu_B^2 = -4\pi^2 \left(\lambda^2 + \frac{\lambda_6^b}{8\pi^2}\right) \left(\int \pdnvol[3] \frac1{\eps(p) - \Sigma} \right)^2,
}
and so the (automatically renormalized) integral can be expressed as
\bel{\label{origin of pm}
  \int \pdnvol[3] \frac1{\eps(p) - \Sigma} = \pm \frac1{2\pi}\sqrt{\frac{\beta^{-2}\mu_B^2 - \sigma}{\lambda^2 + \lambda_6^b/8\pi^2}},
}
where the sign will be chosen to extremize $F$. Putting everything together and using $V_3 = \beta V_2$, we find that the quantity that determines the free energy of both bosonic theories is
\algnl{\notag
   F_{B\pm}\left(\sigma, \mu_B^2, \avg{a}\right)
   &= \frac {NV_2T^2}{2\pi} \Bigg\{\frac{\mu_B^3}3 \pm \frac23 \frac{\left(\mu^2_B - \sigma \beta^2\right)^{3/2}}{\sqrt{\lambda^2 + \lambda_6^b/8\pi^2}} + \\\label{F scalar int}
   &\hspace{-1em} + \frac1{\pi |\lambda|} \int_{\mu_B}^\infty \d y\, y\left[\trm{Im}\, \trm{Li}_2\left(-e^{-y + i\pi|\lambda| + i \avg a} \right) + \trm{Im}\, \trm{Li}_2\left(-e^{-y + i\pi|\lambda| - i \avg a} \right) \right] \Bigg\},
}
extremized with respect to $\sigma$, $\mu_B^2$, $\avg a$, and the sign in front of the second term.\footnote{A slight subtlety must be taken into account here. It is not enough to simply extremize the free energy w.r.t.~$\mu_B$. One must make sure the remaining Schwinger-Dyson equation is also satisfied. Practically, this amounts to requiring $\frac1{\mu_B} \pder F{\mu_B} = 0$ as opposed to just $\pder F{\mu_B} = 0$. It is of note that this just means that one should extremize w.r.t.~$\Sigma \sim \mu_B^2$ and not $\mu_B$; the two procedures do not give the same saddle points.} The entire difference from the commuting model is contained in two minus signs: the overall minus sign in $F_B$, and the sign of the argument of the dilogarithm. We will refer to functions like the one above as ``off-shell free energies,'' and the parameters w.r.t.~which they should be minimized will be written as their arguments.

We now extremize $F_B$ w.r.t.~the eigenvalue clumping position. Demanding $\del F_B /\del\!\avg a = 0$ is equivalent to
\bel{
  \sum_{m = 1}^\infty (-1)^m \, \left( 1+ m \mu_B \right) \, \frac{e^{-m \mu_B}}{m^3} \frac{\sin m\pi |\lambda|}{\pi|\lambda|} \sin m\!\avg a = 0.
}
The sum on the left hand side can be evaluated explicitly in Mathematica. For any $\lambda < 1$, the only solutions of this equation are $\avg a = 0$ and $\avg a = \pi$. Thus, eigenvalues can clump at these two points only. Given all other couplings and choices of signs, there is always a unique $\avg a$ for which there exist solutions to the gap equation. We will explicitly calculate this below. At present, notice that at both values of $\avg a$ the arguments of the two dilogarithms in $F_B$ are equal, and we hence find a simpler off-shell free energy, namely
\bel{
  F_{B\pm}\left(\sigma, \mu_B^2, \avg{a}\right) = \frac {NV_2T^2}{2\pi} \Bigg\{\frac{\mu_B^3}3 \pm \frac23 \frac{\left(\mu^2_B - \sigma \beta^2\right)^{3/2}}{\sqrt{\lambda^2 + \lambda_6^b/8\pi^2}} + \frac2{\pi \lambda} \int_{\mu_B}^\infty \d y\, y\,\trm{Im}\, \trm{Li}_2\left(-e^{-y + i\pi\lambda + i \avg{a}} \right)\Bigg\}.
}
Also, notice that at $\lambda = 1$, any $\avg a$ is allowed. As we will see, in this ``strong coupling'' limit the free energy goes to zero on all saddles under consideration. This means that the effective potential for the holonomies becomes flat, and the $U(1)$  symmetry of the holonomy is restored.

As a check, we rederive the results for the free (regular, $\lambda = \lambda_6^b = 0$) theory that were obtained by \cite{Anninos:2012ft}. Here we set $\sigma = \lambda_6^b = 0$ and take the limit $\lambda \rar 0^+$ in \eqref{FB}. In this regime, the saddle-point equation $\del F_{B\pm}/\del \mu_B^2 = 0$ implies that $\mu_{B\pm} = 0$, i.e.~the gap is zero at both available saddle points.\footnote{This gap equation has a solution only if $e^{i\avg a}$ is chosen to be equal to minus the sign of the second term. This still leaves two saddles, one for each sign/choice of $\avg a$. At one saddle (the plus sign, or $\avg{a}\^{reg}_{B+} = \pi$), $\mu_{B+} = -\lambda \log \lambda$ at weak coupling, while at the other saddle, at $\avg a\^{reg}_{B-} = 0$, $\mu_{B-} = \lambda \log 2$ in the same limit. The thermal masses both vanish as $\lambda \rar 0$, but the different signs of $\avg a$ make the free energies be different.} The minimal free energy is the one at the ``$-$'' saddle, where clumping occurs at $\avg{a}\^{reg}_{B-} = 0$ and the free energy in this ``free theory'' regime is
\bel{
  F\^{reg}_{B-}\Big|\_{free} = \frac{NV_2 T^2}{2\pi} \lim_{\lambda \rar 0} \frac{2}{\pi \lambda} \int_0^\infty\! \d y\, y\, \trm{Im}\,\trm{Li}_2\left(-e^{-y + \pi i \lambda} \right) = - \frac{NV_2 T^2}{\pi} \int_0^\infty\!\d y\, y\, \log\left(1 + e^{-y} \right).
}
Letting $V_2 = 4\pi$, expanding $\log(1 + e^{-y}) = \sum_{m = 1}^\infty  \frac{(-1)^{m + 1}}m e^{-my}$, and doing the integral --- which is just $\int_0^\infty y e^{-my} \d y = 1/m^2$ --- we find that
\bel{
  F\^{reg}_{B-}\Big|_{V_2 = 4\pi,\, \trm{free}} = 4 N T^2 \sum_{m = 1}^\infty \frac{(-1)^m}{m^3} = -3N T^2 \zeta(3),
}
which is precisely the result \eqref{F DDD} of \cite{Anninos:2012ft}.

We end this subsection by noting that a sum representation similar to the one used above can be obtained for the general off-shell free energy \eqref{FB}. By expanding the dilogarithm, this free energy can be written as
\bel{\label{F scalar sum}
   F_{B\pm}\left(\sigma, \mu_B^2, \avg{a}\right) = \frac {NV_2T^2}{2\pi} \left\{\frac{\mu_B^3}3 \pm \frac23 \frac{\left(\mu^2_B - \sigma \beta^2\right)^{3/2}}{\lambda} + 2 \sum_{m = 1}^\infty e^{-m \mu_B} e^{im(\pi + \avg a)}\frac{1 + m \mu_B}{m^3} \frac{\sin\pi m \lambda}{\pi m \lambda} \right\}.
}
This is rather suggestive; in fact, we \emph{conjecture} that the above expression could have been obtained as a partition function of worldlines of particles with thermal mass $\mu_B$, minimized w.r.t.~this thermal mass, with the nontrivial $\mu_B$-dependence entering through the sum of all knots of the worldlines. In this picture, the sum over $m$ is a ``sum over instantons,'' each instanton being a worldline that wraps the thermal circle $m$ times; the factor $\frac{\sin\pi m\lambda}{\pi m \lambda}$ is presumably related to the value of the Polyakov loop with $m$ windings. The differences from the commuting model are transparent in this framework: there is an overall minus factor due to the different sign of the functional determinant, and there is an additional phase factor for each new winding of the worldline around the $S^1$ due to the antiperiodic boundary conditions and the gauge holonomy. We leave the developing of this framework to future work.

\subsection{Fermionic models}

We now turn to the commuting fermion models. Just as in the bosonic case, there are two models of interest: the regular fermion and the critical (Gross-Neveu) fermion with an added marginal deformation. The protocol is the same as in the previous section, and once again the key technical steps have been performed in \cite{Jain:2012qi}
and \cite{Aharony:2012ns}; we merely adapt these tools to our needs. Having reviewed the essential ideas of the computation in the bosonic case, we will be more expeditious with the fermions.

The free energy, expressed in terms of the self-energy $\Sigma(p)$, is
\bel{\label{F fermion start}
  F_F = NV_3 \int \pdnvol[3] \Tr\_{ferm} \left[\log\big(\epsilon(p) - \Sigma(p)\big) + \frac12 \frac{\Sigma(p)}{\epsilon(p) - \Sigma(p)} \right] - NV_3 \frac{\lambda_6^f}6 \sigma^3,
}
where $\Tr\_{ferm}$ is the trace over the two-dimensional internal fermion space, and $\epsilon(p) \equiv i \gamma^\mu (p_\mu - \delta^3_\mu a_i/\beta) + \sigma\1$ is the fermionic bare energy, where $\1$ is the unit operator on the internal fermion space. The self-energy can be split into components, $\Sigma \equiv i\Sigma_\mu \gamma^\mu + \Sigma_0 \1$; at the saddle point only two of these four components are non-zero, and it is convenient to write them as
\bel{
  \Sigma_0(p) - \sigma \equiv f(\beta p_s) p_s, \quad \Sigma_+(p) \equiv g(\beta p_s) p_+.
}
The functions $f$ and $g$ are found via the Schwinger-Dyson equations, which can be shown to give
\algnl{
  &g(x) - f^2(x) = - \frac{\mu_F^2}{x^2},\\
  &\hspace{-1ex}xf(x) = - \sigma\beta - \lambda\sqrt{x^2 + \mu_F^2} + \frac1\pi \! \left[\trm{Im\,Li_2}\!\left(e^{-\sqrt{x^2 + \mu_F^2} + i\pi \lambda + i \avg a} \right) + \trm{Im\,Li_2}\!\left(e^{-\sqrt{x^2 + \mu_F^2} + i\pi \lambda - i \avg a} \right)\right].
}
The fermionic thermal mass $\mu_F \geq 0$ is determined by requiring that $g(x)$ is regular at small $x$, which holds when $\lim_{x \rar 0} x^2 f^2(x) = \mu_F^2$.

With these definitions, the free energy \eqref{F fermion start} can be evaluated. The log term, upon integration and renormalization, gives
\algnl{\notag
  &\int \pdnvol[3] \Tr\_{ferm}\log\big(\epsilon(p) - \Sigma  \big)\\
  &\qquad= \frac1{2\pi\beta^3} \left\{-\frac{\mu_F^3}3 - \frac1{\pi|\lambda|} \int_{\mu_F}^\infty\!\!\! \d y\, y\left[\trm{Im\,Li_2}\!\left(e^{-y + i\pi |\lambda| + i \avg a} \right) + \trm{Im\,Li_2}\!\left(e^{-y + i\pi |\lambda| - i \avg a} \right)\right] \right\},
}
in complete analogy with the bosonic case. The second term cannot be evaluated using as simple an analogy. Upon doing the Matsubara sum and averaging over the gauge holonomy, this term is found to be
\algnl{\notag
  &\int \pdnvol[3] \Tr\_{ferm} \frac12 \frac{\Sigma(p)}{\epsilon(p) - \Sigma(p)} \\ \notag
  &\qquad = \frac1{8\pi^2\beta^3 |\lambda|} \int_0^\infty \d x\, x\, \frac{\mu_F^2 + x^2 f^2(x) + 2\sigma \beta xf(x)}{\sqrt{x^2 + \mu_F^2}}\times \\
  &\qquad \quad \times \left[ \trm{Im}\log\sinh\frac{\sqrt{x^2 + \mu_F^2} - i\pi |\lambda| + i\avg a}2 +\trm{Im}\log\sinh\frac{\sqrt{x^2 + \mu_F^2} - i\pi |\lambda| - i\avg a}2 \right].
}
We may drop all the absolute values. This is going to be self-consistent once $\avg a$ is put on-shell, and at this point it is a necessary step; after it is done, using the second Schwinger-Dyson equation, the integrand can be shown to equal \cite{Aharony:2012ns}
\bel{
  \left(\mu_F^2 + x^2 f^2(x) + 2\sigma \beta xf(x) \right) \frac\pi2 \pder{(xf(x))}x.
}
In other words, this integrand is a total derivative, and with this valuable insight it is possible to regulate the integral above and retain only the cutoff-independent terms (taking into account that $\lim_{x \rar 0} xf(x) = \pm \mu_F$), getting the final answer for the free energy to be
\algnl{\notag
  F_{F\pm}\left(\sigma, \mu_B^2, \avg{a}\right)
  &= -\frac{N V_2 T^2}{2\pi} \Biggr\{\frac{\mu_F^3}3\frac{\lambda \pm 1}{\lambda} + \frac{\sigma \beta \mu_F^2}{2\lambda} - \frac{(\sigma\beta)^3}{6\lambda} + \frac{\pi \lambda_6^f(\sigma \beta)^3}{3} + \\
  &\qquad + \frac1{\pi \lambda} \int_{\mu_F}^\infty \d y\, y\left[\trm{Im\,Li_2}\!\left(e^{-y + i\pi \lambda + i \avg a} \right) + \trm{Im\,Li_2}\!\left(e^{-y + i\pi \lambda - i \avg a} \right)\right] \Biggr\}.
}
Again, the commuting case differs from the original, anticommuting case only by an overall minus sign and by a minus sign in the dilogarithm. In addition, the equation for the thermal mass, $\lim_{x \rar 0} xf(x) = \pm \mu_F$, turns out to be equivalent to $\del F/\del \mu^2_F = 0$, as it should be. Both of these observations fit into our conjecture about the possible alternative derivation of these free energies by summing over particle worldlines and minimizing the sum w.r.t.~the particles' self-energy.

As before, extremizing the free energy w.r.t.~the clumping point we find that only $\avg a= 0$ and $\avg a = \pi$ are allowed. So the off-shell free energy can be written as
\algnl{\notag
  F_{F\pm}\left(\sigma, \mu_B^2, \avg{a}\right)
  = -\frac{N V_2 T^2}{2\pi} \Biggr\{\frac{\mu_F^3}3 &\frac{\lambda \pm 1}{\lambda} + \frac{\sigma \beta \mu_F^2}{2\lambda} - \frac{(\sigma\beta)^3}{6\lambda} + \frac{\pi \lambda_6^f(\sigma \beta)^3}{3} + \\
  &\qquad\qquad\qquad + \frac2{\pi \lambda} \int_{\mu_F}^\infty \d y\, y \,\trm{Im\,Li_2}\!\left(e^{-y + i\pi \lambda + i \avg a} \right) \Biggr\}.
}

As an example, let us examine the regular fermionic theory. In this case we set $\sigma = 0$ and the $\lambda_6^f$-dependence drops out. For the gap equation to have solutions, the sign of $e^{i\avg a}$ must be chosen to be equal to the $\pm$ sign  in the free energy. Once this is done, there are only two saddle points left, and the dominant one must be chosen.  In the case of the free regular fermion ($\lambda \rar 0$), the thermal mass goes to zero as $\mu_F = - \lambda\log\lambda$ at the dominant saddle point (the one with $\avg a\^{reg}_{F+} = 0$), where the free energy is
\bel{
  F_{F+}\^{reg}\Big|\_{free} = - \frac{N V_2 T^2}{\pi} \zeta(3).
}
It is of note that, just like in the ordinary matter case, the free energy of free fermions at high temperatures remains distinct from the free energy of free bosons.

\end{document}